\documentclass[showkeys,nofootinbib,aps,prd,reprint,superscriptaddress,showpacs,floatfix]{revtex4-1}

\usepackage[english]{babel} 
\usepackage{setspace} 

\usepackage[utf8]{inputenc}

\usepackage{hyperref}

\hypersetup{
  pdflang=en, 
  pdfstartview=Fit,
  plainpages=false, 
  bookmarksnumbered=true,
  linkcolor=black,
  filecolor=black,                
  urlcolor=black,                    
  citecolor=black       
}

\usepackage[T1]{fontenc}
\usepackage{lmodern}
\usepackage{soul}
\usepackage{enumitem}
\usepackage{xcolor}
\usepackage[titletoc]{appendix}
\usepackage{overpic}
\usepackage[normalem]{ulem}
\usepackage[nointegrals]{wasysym}

\usepackage[caption=false]{subfig}
\usepackage{graphicx}
\usepackage{nicefrac}
\usepackage{tabularx}
\usepackage{booktabs}
\usepackage{multirow}
\usepackage{amssymb,amsmath}
\usepackage{footnote}
\usepackage{bm}
\usepackage[capitalize]{cleveref}[2012/02/15]

\usepackage{accents}

\usepackage{tabu}

\usepackage{siunitx}
\DeclareSIUnit\parsec{pc}
\DeclareSIUnit\lightyear{ly}

\let\OldAng\ang%
\renewcommand*{\ang}[2][]{%
    \OldAng[scientific-notation=false,#1]{#2}%
}
\DeclareSIUnit\year{yr}
\DeclareSIUnit\erg{erg}
\DeclareSIUnit\msun{M_{\astrosun}}
\DeclareSIUnit{\GeV}{\giga\electronvolt}
\DeclareSIUnit{\TeV}{\tera\electronvolt}
\DeclareSIUnit{\PeV}{\peta\electronvolt}
\DeclareSIUnit{\MeV}{\mega\electronvolt}
\DeclareSIUnit{\eV}{\electronvolt}
\DeclareSIUnit{\smm}{\square\metre\second}
\DeclareSIUnit{\smmr}{\metre^{-2}\second^{-1}}
\DeclareSIUnit{\dc}{d.c.}
\DeclareSIUnit{\pe}{p.e.}
\DeclareSIUnit{\nucleon}{nucleon}

\newcommand{\de}{\mathrm{d}}

\definecolor{desyOrange}{RGB}{242,142,0}
\usepackage{csquotes}

\begin{document}


\title{Measurement of the Iron Spectrum in Cosmic Rays by VERITAS}



\author{Archer, A.}
\affiliation{Department of Physics, Washington University, St. Louis, MO 63130, USA}
\author{Benbow, W.}
\affiliation{Fred Lawrence Whipple Observatory, Harvard-Smithsonian Center for Astrophysics, Amado, AZ 85645, USA}
\author{Bird, R.}
\affiliation{Department of Physics and Astronomy, University of California, Los Angeles, CA 90095, USA}
\author{Brose, R.}
\affiliation{Institute of Physics and Astronomy, University of Potsdam, 14476 Potsdam-Golm, Germany}
\affiliation{DESY, Platanenallee 6, 15738 Zeuthen, Germany}
\author{Buchovecky, M.}
\affiliation{Department of Physics and Astronomy, University of California, Los Angeles, CA 90095, USA}
\author{Bugaev, V.}
\affiliation{Department of Physics, Washington University, St. Louis, MO 63130, USA}
\author{Connolly, M.P.}
\affiliation{School of Physics, National University of Ireland Galway, University Road, Galway, Ireland}
\author{Cui, W.}
\affiliation{Department of Physics and Astronomy, Purdue University, West Lafayette, IN 47907, USA}
\affiliation{Department of Physics and Center for Astrophysics, Tsinghua University, Beijing 100084, China.}
\author{Daniel, M.K.}
\affiliation{Fred Lawrence Whipple Observatory, Harvard-Smithsonian Center for Astrophysics, Amado, AZ 85645, USA}
\author{Falcone, A.}
\affiliation{Department of Astronomy and Astrophysics, 525 Davey Lab, Pennsylvania State University, University Park, PA 16802, USA}
\author{Feng, Q.}
\affiliation{Physics Department, McGill University, Montreal, QC H3A 2T8, Canada}
\author{Finley, J.P.}
\affiliation{Department of Physics and Astronomy, Purdue University, West Lafayette, IN 47907, USA}
\author{Fleischhack, H.}
\email[]{hfleisch@mtu.edu}
\affiliation{DESY, Platanenallee 6, 15738 Zeuthen, Germany}
\affiliation{Michigan Technological University, 1400 Townsend Dr, Houghton, MI 49931, USA}
\author{Fortson, L.}
\affiliation{School of Physics and Astronomy, University of Minnesota, Minneapolis, MN 55455, USA}
\author{Furniss, A.}
\affiliation{Department of Physics, California State University - East Bay, Hayward, CA 94542, USA}
\author{Hanna, D.}
\affiliation{Physics Department, McGill University, Montreal, QC H3A 2T8, Canada}
\author{Hervet, O.}
\affiliation{Santa Cruz Institute for Particle Physics and Department of Physics, University of California, Santa Cruz, CA 95064, USA}
\author{Holder, J.}
\affiliation{Department of Physics and Astronomy and the Bartol Research Institute, University of Delaware, Newark, DE 19716, USA}
\author{Hughes, G.}
\affiliation{Fred Lawrence Whipple Observatory, Harvard-Smithsonian Center for Astrophysics, Amado, AZ 85645, USA}
\author{Humensky, T.B.}
\affiliation{Physics Department, Columbia University, New York, NY 10027, USA}
\author{H\"utten, M.}
\affiliation{DESY, Platanenallee 6, 15738 Zeuthen, Germany}
\author{Johnson, C.A.}
\affiliation{Santa Cruz Institute for Particle Physics and Department of Physics, University of California, Santa Cruz, CA 95064, USA}
\author{Kaaret, P.}
\affiliation{Department of Physics and Astronomy, University of Iowa, Van Allen Hall, Iowa City, IA 52242, USA}
\author{Kelley-Hoskins, N.}
\affiliation{DESY, Platanenallee 6, 15738 Zeuthen, Germany}
\author{Kieda, D.}
\affiliation{Department of Physics and Astronomy, University of Utah, Salt Lake City, UT 84112, USA}
\author{Krause, M.}
\affiliation{DESY, Platanenallee 6, 15738 Zeuthen, Germany}
\author{Krennrich, F.}
\affiliation{Department of Physics and Astronomy, Iowa State University, Ames, IA 50011, USA}
\author{Kumar, S.}
\affiliation{Department of Physics and Astronomy and the Bartol Research Institute, University of Delaware, Newark, DE 19716, USA}
\author{Lang, M.J.}
\affiliation{School of Physics, National University of Ireland Galway, University Road, Galway, Ireland}
\author{Maier, G.}
\affiliation{DESY, Platanenallee 6, 15738 Zeuthen, Germany}
\author{McArthur, S.}
\affiliation{Department of Physics and Astronomy, Purdue University, West Lafayette, IN 47907, USA}
\author{Moriarty, P.}
\affiliation{School of Physics, National University of Ireland Galway, University Road, Galway, Ireland}
\author{Mukherjee, R.}
\affiliation{Department of Physics and Astronomy, Barnard College, Columbia University, NY 10027, USA}
\author{Nieto, D.}
\affiliation{Physics Department, Columbia University, New York, NY 10027, USA}
\author{O'Brien, S.}
\affiliation{School of Physics, University College Dublin, Belfield, Dublin 4, Ireland}
\author{Ong, R.A.}
\affiliation{Department of Physics and Astronomy, University of California, Los Angeles, CA 90095, USA}
\author{Otte, A.N.}
\affiliation{School of Physics and Center for Relativistic Astrophysics, Georgia Institute of Technology, 837 State Street NW, Atlanta, GA 30332-0430}
\author{Park, N.}
\affiliation{Enrico Fermi Institute, University of Chicago, Chicago, IL 60637, USA}
\author{Petrashyk, A.}
\affiliation{Physics Department, Columbia University, New York, NY 10027, USA}
\author{Pohl, M.}
\affiliation{Institute of Physics and Astronomy, University of Potsdam, 14476 Potsdam-Golm, Germany}
\affiliation{DESY, Platanenallee 6, 15738 Zeuthen, Germany}
\author{Popkow, A.}
\affiliation{Department of Physics and Astronomy, University of California, Los Angeles, CA 90095, USA}
\author{Pueschel, E.}
\affiliation{DESY, Platanenallee 6, 15738 Zeuthen, Germany}
\author{Quinn, J.}
\affiliation{School of Physics, University College Dublin, Belfield, Dublin 4, Ireland}
\author{Ragan, K.}
\affiliation{Physics Department, McGill University, Montreal, QC H3A 2T8, Canada}
\author{Reynolds, P.T.}
\affiliation{Department of Physical Sciences, Cork Institute of Technology, Bishopstown, Cork, Ireland}
\author{Richards, G.,T.}
\affiliation{School of Physics and Center for Relativistic Astrophysics, Georgia Institute of Technology, 837 State Street NW, Atlanta, GA 30332-0430}
\author{Roache, E.}
\affiliation{Fred Lawrence Whipple Observatory, Harvard-Smithsonian Center for Astrophysics, Amado, AZ 85645, USA}
\author{Rulten, C.}
\affiliation{School of Physics and Astronomy, University of Minnesota, Minneapolis, MN 55455, USA}
\author{Sadeh, I.}
\affiliation{DESY, Platanenallee 6, 15738 Zeuthen, Germany}
\author{Tyler, J.}
\affiliation{Physics Department, McGill University, Montreal, QC H3A 2T8, Canada}
\author{Wakely, S.P.}
\affiliation{Enrico Fermi Institute, University of Chicago, Chicago, IL 60637, USA}
\author{Weiner, O.M.}
\affiliation{Physics Department, Columbia University, New York, NY 10027, USA}
\author{Wilcox, P.}
\affiliation{Department of Physics and Astronomy, University of Iowa, Van Allen Hall, Iowa City, IA 52242, USA}
\author{Wilhelm, A.}
\affiliation{Institute of Physics and Astronomy, University of Potsdam, 14476 Potsdam-Golm, Germany}
\affiliation{DESY, Platanenallee 6, 15738 Zeuthen, Germany}
\author{Williams, D.A.}
\affiliation{Santa Cruz Institute for Particle Physics and Department of Physics, University of California, Santa Cruz, CA 95064, USA}
\author{Wissel, S.A.}
\affiliation{Physics Department, California Polytechnic State University, San Luis Obispo, CA 94307, USA}
\affiliation{Enrico Fermi Institute, University of Chicago, Chicago, IL 60637, USA}
\author{Zitzer, B.}
\affiliation{Physics Department, McGill University, Montreal, QC H3A 2T8, Canada}

\collaboration{VERITAS collaboration}
\noaffiliation

\date{\today}

\begin{abstract}

We present a new measurement of the energy spectrum of iron nuclei in cosmic rays from \SIrange{20}{500}{\TeV}. The measurement makes use of a template-based analysis method, which, for the first time, is applied to the energy reconstruction of iron-induced air showers recorded by the VERITAS array of imaging atmospheric Cherenkov telescopes. The event selection makes use of the direct Cherenkov light which is emitted by charged particles before the first interaction, as well as other parameters related to the shape of the recorded air shower images. The measured spectrum is well described by a power law $\frac{\de F}{\de E}=f_0\cdot \left(\frac{E}{E_0}\right)^{-\gamma}$ over the full energy range, with $\gamma = 2.82 \pm 0.30 \mathrm{(stat.)} ^{+0.24}_{-0.27} \mathrm{(syst.)}$ and $f_0 = \left( 4.82 \pm 0.98 \mathrm{(stat.)}^{+2.12}_{-2.70} \mathrm{(syst.)} \right)\cdot \SI[scientific-notation=true, exponent-to-prefix=false,retain-unity-mantissa = false]{1e-7} {\per\metre\squared\per\second\per\TeV\per\steradian}$ at $E_0=\SI{50}{\TeV}$, with no indication of a cutoff or spectral break. The measured differential flux is compatible with previous results, with improved statistical uncertainty at the highest energies.

\end{abstract}


\pacs{95.85.Ry,96.50.S-,96.50.sb,96.50.sd}
\keywords{Astroparticle,cosmic rays,cosmic ray iron, IACTs}

\maketitle

\section{Introduction}
\subsection{Cosmic Rays}
More than a hundred years ago, Victor Hess detected the presence of ionizing radiation of extra-terrestrial origin in the atmosphere \cite{ViktorHess}. Since then, the composition and the spectra of these \emph{cosmic rays} have been measured with increasing precision and over an increased energy range. Cosmic rays are mainly composed of protons and fully ionized nuclei, with a small contribution from electrons, positrons, and anti-protons. All known stable elements up to uranium have been detected in cosmic rays (see for example \cite{Hoerandel:2002yg} and references therein). 

The all-particle energy spectrum follows a power-law shape over many orders of magnitude. More precise measurements have revealed several features: a steepening (the \emph{knee}) at around \SI{4}{\peta\eV} \cite{Kulikov_Khristiansen_1958,Hoerandel:2002yg} and a possible second \emph{knee} at around \SI{400}{\peta\eV} \cite{1992JPhG...18..423N,1994ApJ...424..491B}, a subsequent flattening (the \emph{ankle}) above about \SI{4}{\exa\eV} \cite{Horandel:2006jd} as well as a final cutoff at around \SI{40e18}{\eV} \cite{Settimo:2012zz}.


It is currently assumed that cosmic rays below the knee are accelerated within our own galaxy. The main class of source candidates is assumed to be young to middle-aged supernova remnants (SNRs), which have a sufficient energy budget to produce the detected fluxes of cosmic rays \cite{1989A&A...225..179D}, and several of which have been shown to accelerate hadronic cosmic rays \cite{Ackermann807}. However, it is not clear whether SNRs are able to accelerate cosmic rays up to the knee, or whether another class of sources is needed to explain the Galactic cosmic ray spectrum \cite{0954-3899-31-5-R02}. Precision measurements of the elemental spectra, or equivalently of the energy-dependent composition, are needed to find features in the energy spectra and to disentangle their origins. 

The elemental energy spectra also follow power-law shapes over many orders of magnitude, with the caveat that composition measurements are challenging at \si{\peta\eV} energies and above. The \emph{knee} in the all-particle spectrum may be the result of a rigidity-dependent cutoff in the single-element spectra, see for example \cite{Gaisser:2013bla}.

In addition to the features at \si{PeV} energies and above, measurements by PAMELA, recently confirmed by AMS-02, have found a slight, but significant spectral hardening in proton and helium spectra above a few hundred \si{GeV} \cite*{2011Sci...332...69A,PhysRevLett.114.171103,PhysRevLett.115.211101}. Similarly, the CREAM experiment has seen an indication of spectral hardening in heavy elements (carbon to iron) above roughly \SI{200}{\GeV\per\nucleon} \citep{2010ApJ...714L..89A}. The underlying cause of the hardening is unclear; possible explanations include effects due to nearby sources \cite{refId0}, effects related to the presence of multiple strong shocks in the accelerating SNRs \cite{0004-637X-763-1-47}, and propagation effects in the Galaxy \cite{2041-8205-752-1-L13}.  

Up to some tens or even hundreds of \si{\TeV}, the energy, momentum, charge, and mass of incident cosmic rays can be measured directly by balloon-borne or space-based detectors (see \cite{PDG2014} and references therein for an overview of results). These typically have a charge resolution of better than one electron charge, enabling them to clearly separate the elements, but are limited by low statistics at high energies, especially for heavy elements (see for example \cite*{0004-637X-707-1-593,2011ApJ...742...14O,2010ApJ...714L..89A}). Above hundreds of \si{\TeV}, cosmic rays are best detected by extensive-air-shower arrays. These arrays detect part of the \emph{shower} or cascade of particles which is the product of the interaction of an incident cosmic ray with a nucleus in the atmosphere. The primary energy and (to some extent) the charge or mass of the incoming nucleus can be inferred from the distribution of secondary particles (mostly electrons and muons) on the ground. Typically, air-shower experiments have worse charge resolution compared to direct detection experiments, but their larger collection area lets them accumulate more statistics. 

\subsection{The Iron Spectrum}
\label{sec:ironothers}
After protons and helium, which make up the overwhelming majority of cosmic rays, iron is the third most abundant element in cosmic rays at \si{\TeV} energies. According to \citep{Hoerandel:2002yg}, cosmic rays at \SI{1}{\TeV} per nucleus are composed of roughly \SI{38}{\percent} hydrogen, \SI{25}{\percent} helium, and \SI{9}{\percent} iron. The iron spectrum is of particular interest to decide whether features in the spectrum are proportional to the elemental charge or mass, which can provide a clue as to their origins. The iron spectrum has been measured over a wide energy range (\SIrange{50}{1e9}{\GeV}, see for example \cref{fig:ironspectrum}); however, the range from \SIrange{10}{1000}{\TeV} is not well covered by either direct detection methods or air shower arrays because the former have limited statistics and the latter do not have good charge resolution at these energies. 

Several balloon-borne detectors have been able to measure the spectra of heavy elements in the \si{\TeV} region. For example, the CREAM collaboration measured the iron spectrum from \SI{1}{\TeV} to about \SI{100}{\TeV}, with a charge resolution of $0.5e$ and an energy resolution of about \SI{30}{\percent} \citep{0004-637X-707-1-593}. They found the spectrum to be compatible with a power law with index $\gamma=2.63 \pm 0.11$. Their measurement is statistics-limited at the highest energies: The highest-energy data point ($E>\SI{50}{\TeV}$) contains only 4 events. However, due to the good charge resolution, the sample is virtually background-free. The TRACER collaboration has published measurements taken during two different long-duration flights, with a charge resolution of about $0.6e$ \cite*{Ave:2008as,2011ApJ...742...14O}. The energy resolution was improved in the second flight (\SI{30}{\percent} vs \SI{40}{\percent} in the first flight for a \SI{5}{\TeV} iron nucleus). Their measurements were compatible with a power-law spectrum as well, with an index of $2.68\pm0.04$. Again, their results are statistics-limited at the highest energies, with only eleven events in the highest energy bin (above \SI{50}{\TeV}) in the first, longer duration flight. 

Imaging atmospheric Cherenkov telescopes (IACTs) are able to `bridge the gap' between direct detection and air shower measurements and provide improved measurements of the iron spectrum in this energy range. Rather than relying on parts of the air shower to reach the ground, they detect Cherenkov emission from the charged shower components, and thus can measure air showers with energies above a few hundred \si{\GeV}. IACTs are sensitive to the elemental charge of primary particles with \si{\TeV} energies or higher by exploiting the \emph{direct} Cherenkov (DC) radiation which is emitted by charged primaries prior to the first interaction \citep{Kieda}; its intensity is proportional to the square of the elemental charge.

Both H.E.S.S. and VERITAS have performed measurements of the energy spectrum of iron-like elements in  cosmic rays, using the presence of DC light as a selection criterion \cite*{hess,wissel}. H.E.S.S. measured the iron spectrum in the energy range from \SIrange{13}{200}{\TeV}, and VERITAS did the same for the energy range from \SIrange{20}{140}{\TeV}. The results agree with each other and are compatible with a power-law spectrum over the respective energy ranges (see \cref{fig:ironspectrum}). 

In this paper, we present an updated measurement of the iron spectrum made with the VERITAS experiment. We present a template likelihood method which, for the first time, is adapted for the reconstruction of iron-induced showers and which improves the reconstruction of both the energy and the event geometry (arrival direction and core position). With this analysis, we were able to extend the measured spectrum up to \SI{500}{\TeV}.

\subsection{Direct Cherenkov Emission}
IACTs detect air showers via the Cherenkov light emitted by their charged componenents as they propagate throught the atmosphere. In contrast to photons, cosmic rays with sufficient energy emit Cherenkov light themselves prior to interacting with air nuclei. This \emph{direct} Cherenkov (DC) emission is emitted coherently by the primary nucleus, and its intensity is proportional to the square of the charge of the cosmic-ray particle. Hence, the presence of DC light is a signature of showers induced by heavy elements, of which iron is the most abundant. 

The density of the atmosphere decreases with increasing altitude. Thus, there is a maximum altitude for Cherenkov emission, depending on the particle's speed. For example, an iron nucleus with an energy of \SI{20}{\TeV} approaching the VERITAS site from overhead has a chance of roughly \SI{50}{\percent} to start an air shower before it reaches its Cherenkov threshold (about \SI{37}{\kilo\metre} above sea level). Depending on the instrument sensitivity to DC light and on further analysis cuts, this implies an energy threshold of about \SIrange{10}{20}{\TeV} for an analysis using DC light to identify iron-induced showers. 

Compared to the Cherenkov light emitted by air showers, which forms an extended image in the VERITAS camera, the DC light is concentrated within one or two pixels. The contribution from DC light to a shower image can thus be identified by selecting images with one unusually bright pixel close to the reconstructed arrival direction of the primary particle (cf. \cref{fig:template,fig:ironincamera}).

\section{Instrument and Data Selection}
\subsection{VERITAS}
VERITAS\footnote{\protect\url{http://veritas.sao.arizona.edu}}
\cite*{veritas,naheeperformance} is an array of four IACTs located in southern Arizona in the USA. Each telescope has a mirror area of about \SI{100}{\metre\squared}, consisting of 350 hexagonal facets in a Davies-Cotton design \cite{1957SoEn....1...16D} with a focal length of \SI{12}{\metre}. Each has a camera comprising 499 photomultiplier tubes (PMTs) or `pixels' and covering a field-of-view of approximately \ang{3.5}. Each PMT has a field of view of about \ang{0.14} diameter. 

VERITAS has been upgraded several times. The array layout was changed in 2009, improving the effective collection area and the angular reconstruction. In 2012, the PMTs were replaced with higher quantum efficiency models to improve the response of the array in particular to low-energy showers. In its current configuration, VERITAS is sensitive to gamma-ray induced showers from \SI{80}{\GeV} to greater than \SI{30}{\TeV}. 

The VERITAS readout system uses FADCs with a sampling time of 2ns, allowing the charge integration time to be chosen at the analysis stage. The conversion factor between the integrated recorded charge measured in digital counts (d.c.) and the number of photo-electrons (p.e.) at the photocathode depends on the choice of integration window. For this analysis, an integration window of six samples was used, resulting in an effective conversion factor of \SI[per-mode=symbol]{3.3}{\dc\per\pe}

Even in the absence of a Cherenkov pulse, the FADCs record a non-constant signal due to night-sky background light (NSB) and electronics noise. This so-called \emph{pedestal level} and its variance are measured constantly during data-taking. 

\subsection{Data Selection}
\label{sec:dataselection}
VERITAS was designed to record showers initiated by VHE (very high energy, $E>\SI{100}{\GeV}$) gamma rays. However, the event rate is actually dominated by showers initiated by cosmic rays (including iron nuclei). In this work, we make use of this cosmic-ray background to measure the iron spectrum. No changes to the observation procedures are required; existing data from observations of gamma-ray sources and source candidates are used and analyzed using a special analysis chain.

VERITAS data were selected according to the following requirements:

\begin{enumerate}
\item Detector configuration: Data runs recorded between 2009 September and 2012 May, a period in which no major upgrades were performed.
\item Observing season: Data runs recorded in winter-like months (October to March). (Most of the VERITAS data are recorded in the winter months; adding data recorded in summer-like months would have required more simulations with a different atmospheric profile without a large improvement in statistical uncertainty.) 
\item Data quality: Data runs taken under nominal observing conditions, with all four telescopes active, and under clear, moonless skies.
\item Elevation: Average elevation above \ang{80}.
\end{enumerate}

210 data runs passed all criteria, corresponding to \SI{71}{\hour} livetime. The data were calibrated and analyzed following the procedures described in \cref{sec:methods}.

\section{Analysis Methods}
\label{sec:methods}
\subsection{Geometric Reconstruction}
The data analysis methods used for analysis of gamma-ray data recorded by VERITAS are described in \cite{veritasanalysis}. After pedestal subtraction and image cleaning, the Hillas parameters \citep{hillas} are calculated for each image. The arrival direction of the primary particle as well as the shower core position are determined from the positions and orientations of the images in the cameras (only events with images in two or more cameras are considered). Look-up tables are used to determine the primary energy, given the total signal (\emph{size}) and the impact distance between the detector and the shower core.

Look-up tables are also used to determine the mean reduced scaled width ($MSCW$) and mean reduced scaled length ($MSCL$) for events with images in $N$ telescopes, used in gamma/hadron separation:  

\begin{align}
MSCW = \frac{ \sum\limits_{i=1}^N  MSCWT_i \cdot c_i }{ \sum\limits_{i=1}^N  c_i} 
\end{align}
with weights $c_i = \left( \frac{ \left\langle w_{MC}(s_i,D_i)\right\rangle }{\sigma_{MC}^{width}(s_i,D_i)} \right)^2$, the reduced scaled width per image $MSCWT_i = \frac{w_i - \left\langle w_{MC}(s_i,D_i)\right\rangle }{\sigma_{MC}^{width}(s_i,D_i)}$; $w_i$, $s_i$, and $D_i$ the \emph{width}, \emph{size}, and impact distance of the $i$th image/telescope; and $\left\langle w_{MC}(s,D)\right\rangle$ and $\sigma_{MC}^{width}(s,D)$ the median and \SI{90}{\percent} containment interval of the distribution of the \emph{width} parameter in simulations for a given \emph{size} $s$ and impact distance $D$. The mean reduced scaled length is defined analogously.

To reconstruct iron-induced showers, a similar geometric analysis was performed, albeit with tighter quality cuts (requiring at least 70 hit pixels in each of the four cameras, in contrast to the threshold of four hit pixels commonly used for the reconstruction of gamma-ray induced showers). In a typical run, only \SIrange{1}{3}{percent} of the recorded events pass this requirement. Dedicated look-up tables were produced for the energy and mean reduced scaled parameters for iron showers. The energy, arrival direction, and core position were used as the starting values in the likelihood fit and the mean reduced scaled parameters were included in the random forest classifier that was used to estimate the remaining background (cf. \cref{sec:rf}).

\subsection{Template Likelihood Reconstruction}
The stereoscopic reconstruction described above is very robust for most gamma-ray showers. However, it does have large uncertainties for high-energy showers where only part of the image is contained in the camera, as well as for cosmic-ray showers, whose images are in general less smooth than gamma-ray images. Also, this technique does not use all available information, relying only on  the total signal as well as the mean and variance of the coordinates of the hit pixels. More advanced techniques are needed to obtain optimal performance. One such method is the template likelihood fit, pioneered by the CAT experiment for the reconstruction of gamma-ray induced showers \cite*{cattemplate,Parsons201426,frogs}. A related approach uses a semi-analytic description of the air showers instead of image templates \cite{deNaurois2009231}.  

In this paper, we will show that the template likelihood method can also be adapted to reconstruct and identify cosmic-ray induced showers. The main change that needs to be made is the inclusion of shower-to-shower fluctuations into the likelihood function. These fluctuations are much larger for cosmic-ray induced showers compared to purely electromagnetic showers. The implementation of the template likelihood fit described here is based on the FROGS code\footnote{\url{http://www.physics.utah.edu/gammaray/FROGS/}}, a template likelihood reconstruction code for the analysis of gamma-ray induced showers recorded by VERITAS \cite{frogs}.

The likelihood fit requires a model for the probability distribution of the signal in each camera pixel, depending on some properties of the primary particle, and a likelihood formula comparing the recorded signal in each pixel to the model predictions. For each event to be reconstructed, the event parameters are then varied to maximize that likelihood, given the recorded signal. The set of event parameters which maximizes the value of the likelihood function can be used as an estimator for the true values. An additional feature of the template method is that in addition to reconstructing event parameters such as the energy and direction of the primary particle, the value of the final likelihood can be used as an additional classifier to separate signal and background events, as well as for quality selection. 

In this implementation, the model parameters are:
 
\begin{itemize}
\item the primary particle's energy $E$,
\item the height $h$ of the first interaction of the primary with an atom in the atmosphere,
\item the direction of the primary particle relative to the pointing direction of the telescopes ($X_s$, $Y_s$),
\item the projected position $X_p$, $Y_p$ of the shower core on the ground.
\end{itemize}

This set of model parameters describing a given event is abbreviated as $$\mathbf{\Theta}=(E,h,X_p,Y_p,X_s,Y_s).$$

To first order, the shape and magnitude of the template image depend on only three independent parameters ($E$, $h$, and the impact distance $D$ between the shower core and the telescope). The remaining three parameters define the placement and orientation of the image in the camera. The radial dependence of the optical point spread function is neglected here. This approach is justified since the 68\% containment  radius is always smaller than or comparable to the PMT size.

The image shape and magnitude also depend on the absolute zenith and azimuth angles. Only showers with zenith angles below \ang{15} were considered for this analysis. For these small zenith angles, the zenith-dependence is small (less than a $3.5\%$ increase in  traversed atmospheric material); neglecting it leads to a small bias in the energy reconstruction of order \SI{10}{\percent}, which contributes to the systematic uncertainty (cf. \cref{sec:systunc}). The azimuth dependence (due to  the Earth's magnetic field) was found to be negligibly small. 

\subsubsection{Image Template Generation}
\begin{figure*}[tb]
\flushleft

\subfloat[]
	[Image template for a first interaction height of \SI{33}{\kilo\metre}, an energy of \SI{15}{\TeV} and impact distance of \SI{80}{\metre}. There is a contribution from DC light about \ang{0.2} offset from the primary direction.\label{fig:template-H5-E15-D80.pdf}]
	{\includegraphics[width=0.48\textwidth]{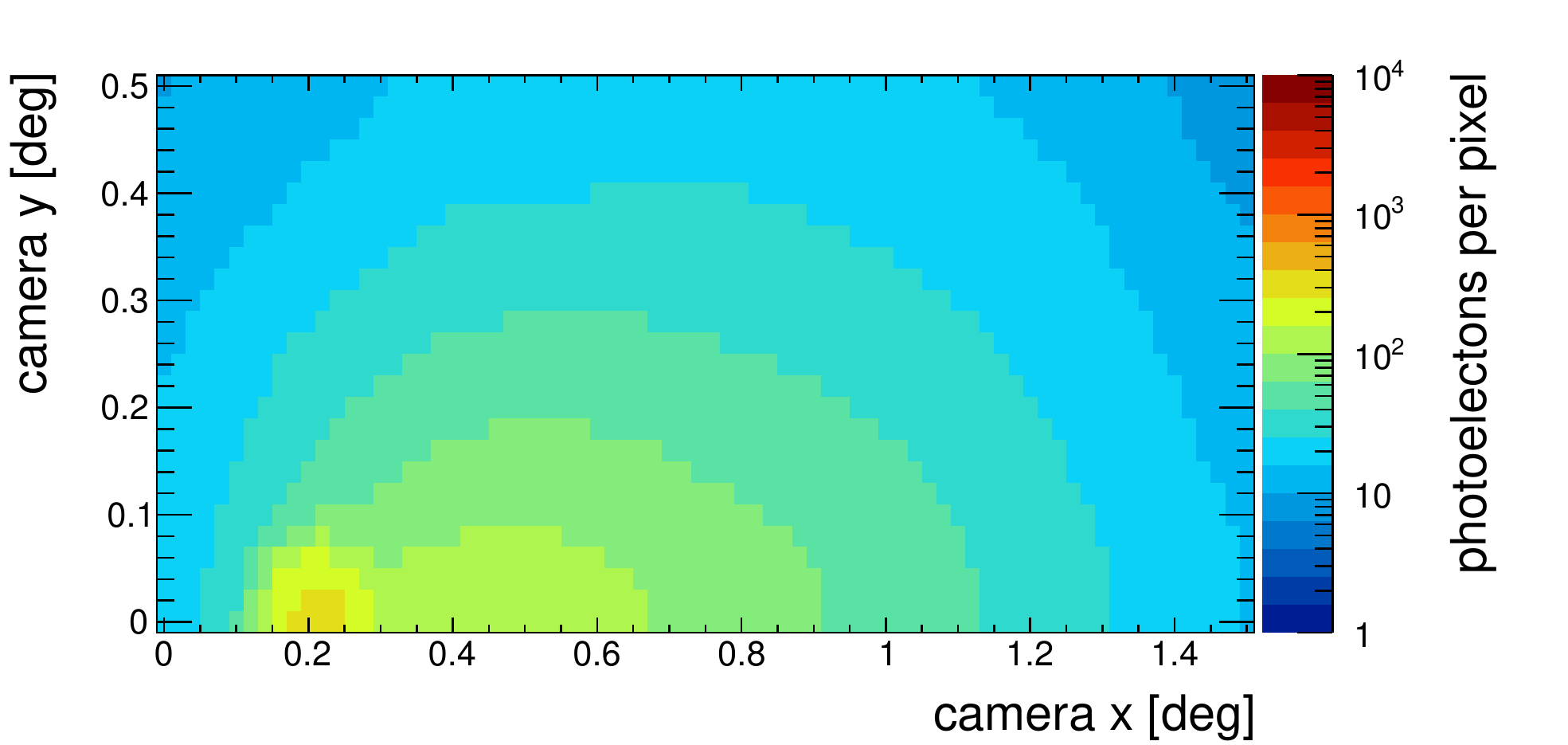}  }\hfill
\subfloat[]
	[Image template for a first interaction height of \SI{33}{\kilo\metre}, an energy of \SI{80}{\TeV} and impact distance of \SI{80}{\metre}. There is a contribution from DC light about \ang{0.2} offset from the primary direction.\label{fig:template-H5-E80-D80.pdf}]
	{\includegraphics[width=0.48\textwidth]{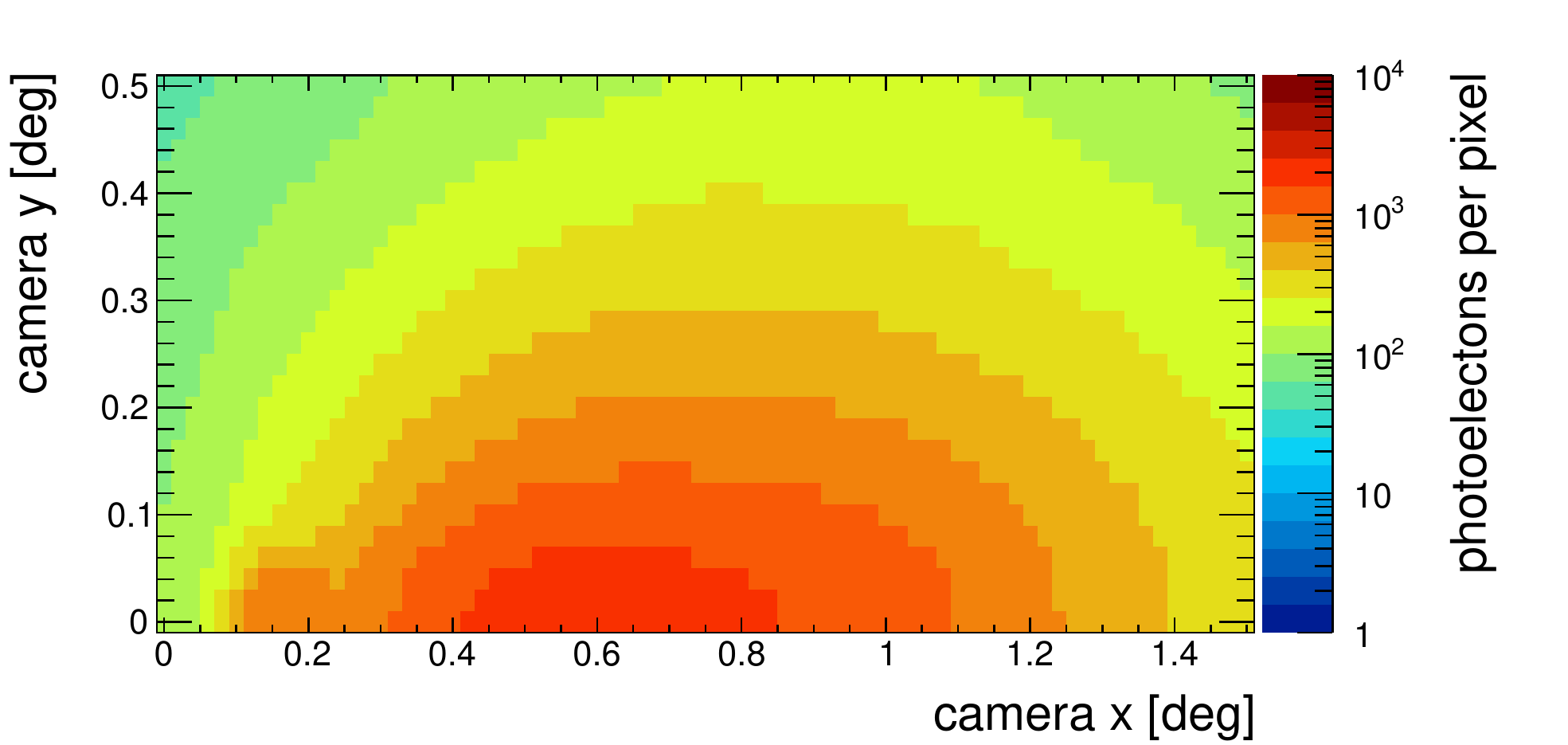}  }

\subfloat[]
	[Image template for a first interaction height of \SI{33}{\kilo\metre}, an energy of \SI{30}{\TeV} and impact distance of \SI{20}{\metre}. The image is nearly round; the contributions from DC emission and the air shower emissions overlap.\label{fig:template-H5-E30-D20.pdf}]
	{\includegraphics[width=0.48\textwidth]{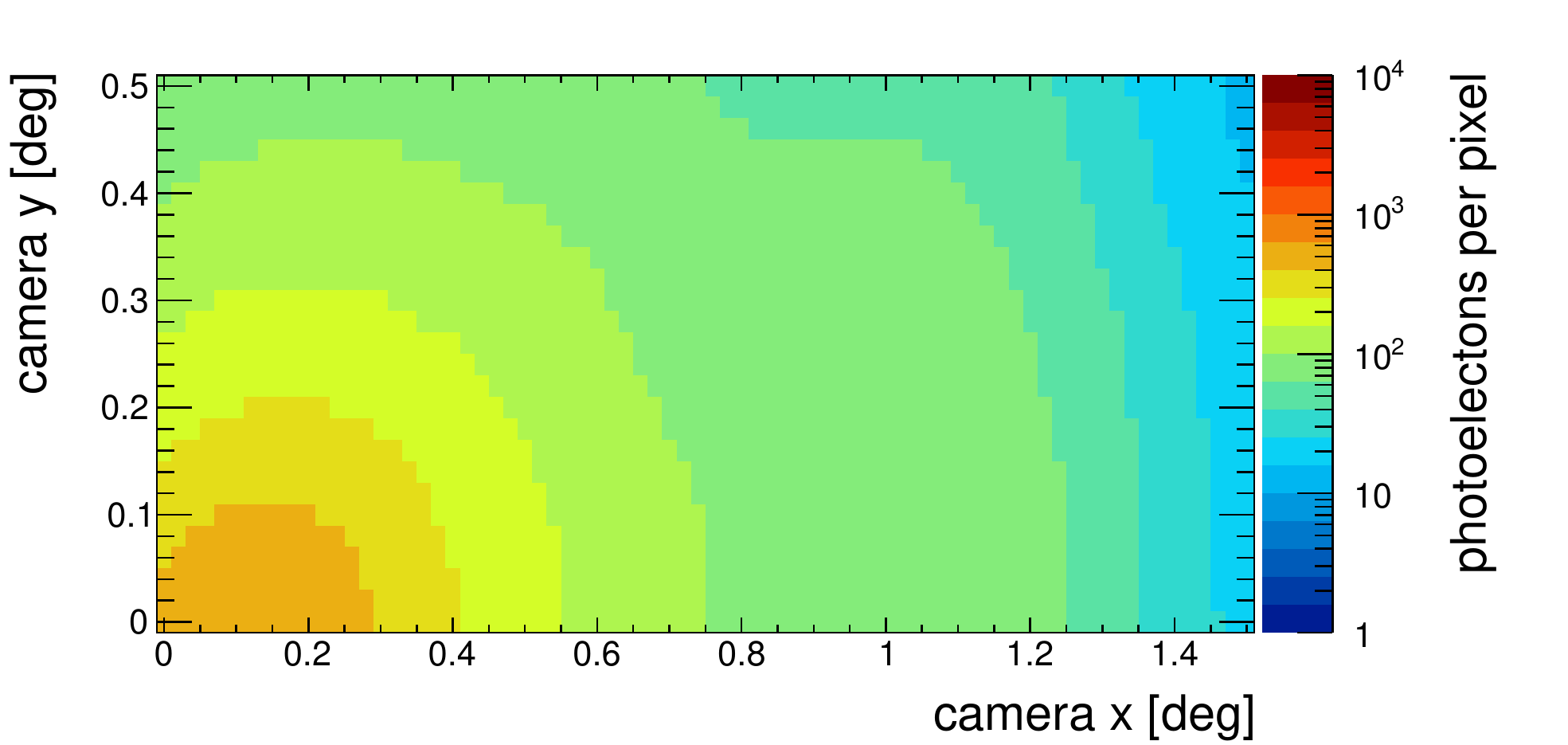}  }\hfill
\subfloat[]
	[Image template for a first interaction height of \SI{33}{\kilo\metre}, an energy of \SI{30}{\TeV} and impact distance of \SI{200}{\metre}. The image is very elongated and there is no contribution from DC light since the impact distance is too large.\label{fig:template-H5-E30-D200.pdf}]
	{\includegraphics[width=0.48\textwidth]{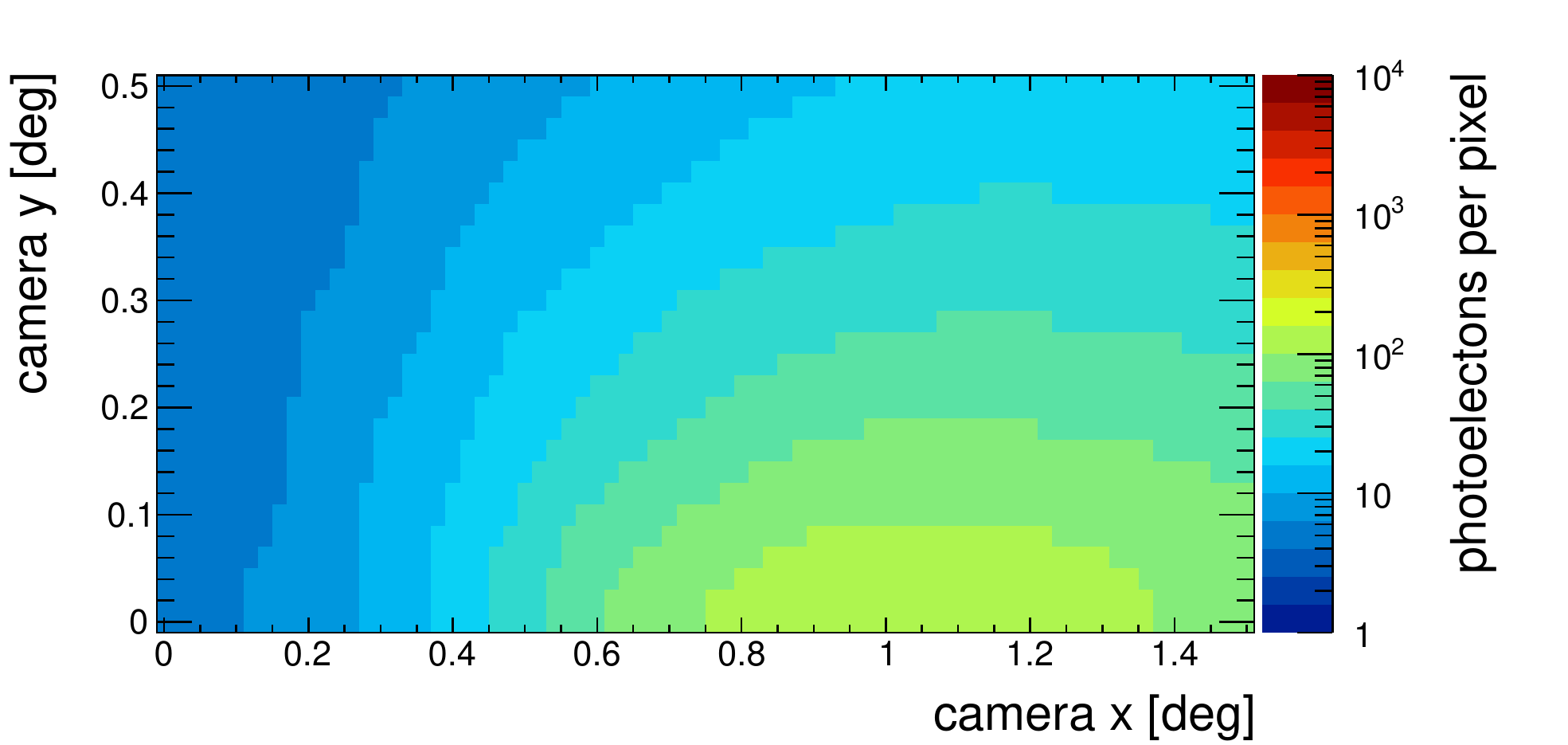}  }

\caption[Iron shower templates]{Iron shower templates (mean number of photo-electrons (p.e.) per camera pixel) for the VERITAS cameras. All templates were produced for zenith angle \ang{0}. The x-axis is chosen to be the symmetry axis of the image, the y-axis is perpendicular to the x-axis so that (0,0) marks the direction of the primary particle. The images are symmetrical about the y axis; only the top half ($y>0$) of the images is plotted. These templates result from averaging several thousand simulated air showers.}
\label{fig:template}
\end{figure*}

\begin{table*}[tb]
\caption[Image template grid settings]{Grid parameters for the generation of the image templates. $D$ corresponds to the impact distance (distance between the shower core and telescope). The values of the first interaction height $h$ are chosen to provide even coverage of the likely first interaction heights. The $i$th height $h_i$ corresponds to column depth $\chi_i=\lambda\cdot\ln\left(\frac{N-i-\Delta}{N} \right)$ with $\lambda=13\,\frac{g}{cm^2}$, $N=11$ and $\Delta=0.5$ for $i>0$, $h_0$ is set to \SI{70}{\kilo\metre}. }
\label{tab:templategrid}
\begin{tabular*}{14cm}{@{\extracolsep{\fill} } lllll}
\toprule
Parameter & No. of steps & Step size & First value & Last value \\ 
\midrule
$\log_{10}\left(\frac{E}{\si{\TeV}}\right)$ & 17 & 0.1 & $\log_{10}(10)$ & $\log_{10}(500)$ \\ 
$D$		& 31 & \SI{10}{\metre}  & \SI{0}{\metre} & \SI{300}{\metre} \\ 
$h$		& 11 & $-$		& \SI{70}{\kilo\metre} & \SI{22}{\kilo\metre} \\
\bottomrule
\end{tabular*} 
\end{table*} 
 
The model for the distribution of Cherenkov light in the camera is obtained using Monte Carlo simulations of iron-induced showers at a fixed `grid' in $E$, $h$, and $D$, see \cref{tab:templategrid}. Showers are simulated originating from zenith (zenith angle $Ze=\ang{0}$), with the telescopes pointing at zenith as well. The number of simulated showers for each grid point depends on the simulated energy; ranging from 10000 showers at the lowest energy (\SI{10}{\TeV}) to 400 showers at the highest energy (\SI{500}{\TeV}).

The \emph{CORSIKA} software \cite{corsika} with the QGSjet II-03 hadronic interaction model \cite*{qgsjet1,qgsjet2} is used for the simulation of the shower development and the emission of Cherenkov light. The GrOptics\footnote{\protect\url{http://otte.gatech.edu/care/}} package is used for ray-tracing in the telescope, giving the distribution of Cherenkov light in the camera plane. All four VERITAS telescopes were built according to the same design specifications. Hence, the same templates were used for all telescopes. A correction factor for shadowing of light by the camera box and telescope structure is applied to the templates. The number of photons is integrated over a circle with diameter \ang{0.14}, corresponding to the size of the pixels in the VERITAS camera. The result is multiplied by the wavelength-dependent mirror reflectivity and the PMT quantum-efficiency and integrated over the emitted wavelengths to obtain the expected number of p.e. per pixel. Light distributions for showers not originating from the camera center are obtained by displacing and rotating the predicted image in the camera. Distributions for arbitrary values of $E$, $h$, and $D$ are obtained by interpolating between the grid points. Some examples for the distribution of photo-electrons in the camera for iron showers can be seen in \cref{fig:template}. 

To be able to compare the template prediction to the measured signal, the predictions are multiplied by the camera pixel's gain, taking into account the signal `lost' due to the finite integration window. Correction factors are applied to account for the measured differences in throughput between the telescopes. The template predictions are adjusted for non-linearities and saturation effects in the PMTs, amplifiers, and readout electronics. 

\subsubsection{Likelihood Fit}
The development of air showers, the propagation of Cherenkov light, and the detector response involve stochastic processes. There are several factors which cause the signal observed in a given pixel to deviate from the average template prediction for an event with event parameters $\mathbf{\Theta}$:

\begin{enumerate}
\item Fluctuations in the number of Cherenkov photons emitted by the air shower (due to fluctuations in the shower development). The resulting probability distribution is assumed to be Gaussian; mean and standard deviation values are taken from simulations. 
\item Fluctuations in the number of photo-electrons per pixel, given the number of emitted Cherenkov photons, e.g., due to the stochastic nature of absorption/scattering in the atmosphere. This follows a Poisson distribution, which can be approximated by a Gaussian for large numbers of photo-electrons.
\item Fluctuations in the recorded signal (in digital counts) for a given input (a certain number of p.e.s). This distribution is Gaussian. The mean is given by the true  number of photo-electrons. The two main contributions to the variance are the pedestal variance, $\sigma_p^2$, and the variance in the PMTs' single-p.e. response, $\sigma_e^2$. The pedestal variance is mostly due to the night sky background and is measured continuously during data-taking. The gain variance is measured in special calibration runs.
\end{enumerate}

In contrast to previous implementations of the template likelihood method \cite*{cattemplate,Parsons201426,frogs}, which only took the second two contributions into account, the shower-to-shower fluctuations cannot be neglected for cosmic-ray induced showers. 

The probability distribution for the signal in one pixel is then given by the convolution of the three components mentioned above, cf. \cite*{deNaurois2009231,Parsons201426,frogs}:

\begin{align}
P(&q|s(\mathbf{\Theta}),\sigma_p, \sigma_e, \sigma_s(\mathbf{\Theta})) = \int \mathrm{d}\mu ~ \overbrace{G(\mu|s(\mathbf{\Theta}), \sigma_s(\mathbf{\Theta}))}^{1.} \nonumber \\ \cdot& \sum\limits_n \overbrace{Poi(n|\mu)}^{2.} \cdot \overbrace{G(q|n,\sqrt{\sigma_p^2 + n\sigma_e^2})}^{3.} \nonumber \\
\approx& ~ G(q|s(\mathbf{\Theta}),\sigma(\mathbf{\Theta})) \nonumber \\ ~&~\mathrm{with} ~~ \sigma = \sqrt{\sigma_p^2+s\left(1+\sigma_e^2\right)+\sigma_s^2(\mathbf{\Theta})}
\label{eq:ppixel}
\end{align} 

Where

\begin{itemize}
\item $\mathbf{\Theta}=(E,h,X_p,Y_p,X_s,Y_s)$ is the (true) set of parameters describing the event,
\item $G(x|\mu,\sigma)= \frac{1}{\sigma\sqrt{2\pi} } \cdot \exp\left( -\frac{(x-\mu)^2}{2\sigma^2} \right)$ is the normal distribution with mean $\mu$ and width $\sigma$,
\item $Poi(n|\mu)=\frac{\mu^n e^{-\mu}}{n!}$ is the Poisson distribution with mean $\mu$,
\item $q$ is the pedestal-corrected integrated charge (converted to units of p.e.),
\item $s(\mathbf{\Theta})$ and $\sigma_s(\mathbf{\Theta})$ are the predicted average and standard deviation of the number of photo-electrons in each pixel,
\item $\mu$ is the predicted number of photo-electrons in a given shower, assumed to follow a Gaussian distribution with mean $s$ and width $\sigma_s$,
\item $n$ is the number of photo-electrons, assumed to follow a Poisson distribution with mean $\mu$,
\item $\sigma_e^2$ is the variance of the signal (integrated charge) from a single photo-electron, and
\item $\sigma_p^2$ is the variance of the FADC pedestal of the given pixel.
\end{itemize}
The Poisson distribution may be approximated by a Gaussian distribution if the predicted number of photo-electrons is large. If the predicted number of photo-electrons is small (which happens at the edges of the images), the fluctuations due to the night-sky background noise dominate the distribution, and so again the Gaussian approximation may be used in the convolution.

The likelihood of the primary particle having certain parameters $\mathbf{\Theta}$, given the measured signal $q_i$ in pixel $i$, is given by the same function:

\begin{align}
L_i(\mathbf{\Theta}|q_i,\sigma_p, \sigma_e) ~=~ P\left(q_i|s_i(\mathbf{\Theta}),\sigma_{p,i}, \sigma_{e,i}, \sigma_{s,i}(\mathbf{\Theta})\right) 
\end{align}

The overall likelihood function for each event is given by the product over all pixels, or equivalently, by the sum of the log-likelihood values:

\begin{align}
-\ln L(\mathbf{\Theta}) =& - \sum\limits_{\mathrm{pixel}~i} \ln L_i(\mathbf{\Theta}) \nonumber \\ =& -\sum\limits_{\mathrm{pixel}~i} \ln P(q_i|s_i(\mathbf{\Theta}),\sigma_i(\mathbf{\Theta})). 
\end{align}

The true event parameters $\mathbf{\Theta}$ are unknown in data. The event parameters $\mathbf{\Theta}=(E,h,X_p,Y_p,X_s,Y_s)$ are adjusted to minimize the negative log-likelihood function $-\ln L$ for each event:

\begin{align*}
\hat{\mathbf{\Theta}} ~\mathrm{such~that}~ \hat{L} = L(\hat{\mathbf{\Theta}}) \geq L(\mathbf{\Theta}) ~\mathrm{for~all}~ \mathbf{\Theta}   
\end{align*}

For this study, the Levenberg-Marquardt algorithm for multi-dimensional minimization, as implemented in the Gnu Scientific Library\footnote{\protect\url{http://www.gnu.org/software/gsl/}}
, was chosen. The maximum-likelihood estimators $\hat{\mathbf{\Theta}}$ obtained in this way are consistent estimators for the true values (cf. e.g. \cite{Behnke:1517556}). 

For debugging or performance evaluation, it can be useful to fix one or several event parameters (e.g., to their true values for simulated events) to exclude them from the fit. In this case, the minimizer returns the best-fit parameters for the given constraints. 

The minimizing algorithm requires a starting point; the results of the standard moment-based event reconstruction are sufficient for this purpose. For the first interaction height, no starting value is available and thus it was attempted to use a fixed value of \SI{33.3}{\kilo\metre}, the approximate median for iron showers in the energy range studied here. However, the fit of the first interaction height was not stable, and there was a large discrepancy in the distribution of the first interaction height when comparing data to simulations. Accordingly, this parameter was excluded from the fit and fixed to the previously mentioned median value. 

\subsubsection{Average Value of the Likelihood and Goodness of Fit}
Following an approach similar to \cite{deNaurois2009231}, the mean and standard deviation of $l=-2\ln L$ (over an ensemble of events with fixed event parameters $\mathbf{\Theta}$ and $N$ pixels) are given by

\begin{align}
\left\langle l \right\rangle_q =& N\cdot \left( 1 + \ln 2\pi\right) \nonumber \\ +& \sum\limits_i \ln\left(\sigma_{p,i}^2 + s_i(\mathbf{\Theta})\cdot\left(1+\sigma_{e,i}^2\right) + \sigma_{s,i}^2(\mathbf{\Theta})\right)
\end{align}

and $\sigma_l = \sqrt{2N}$.

This is the distribution of the measured likelihood values for a fixed set of true parameters. Since the true parameters are unknown for actual measurements, we must instead use the maximum-likelihood estimators, $\hat{\mathbf{\Theta}}$ and $\hat{l}=-\ln \hat{L}$. For the application presented here, $\mathbf{\Theta}$ is over-constrained because the number of pixels is much larger than the number of free parameters. Hence, $\hat{\mathbf{\Theta}}\approx \mathbf{\Theta}$, and it can be assumed that the distributions of $\hat{l}$ and $l$ are similar.

In that case, it is appropriate to define a goodness-of-fit function,

\begin{align}
G=\frac{-2\cdot\ln \hat{L} - \left\langle l \right\rangle_q }{\sqrt{2 N}}.
\end{align}

If $\hat{\mathbf{\Theta}}$ were a perfect estimator for $\mathbf{\Theta}$, the goodness of fit $G$ would have mean $0$ and width $1$. In practice, as $\hat l$ is smaller than $l$ by definition, the mean of $G$ will be slightly below 0.

The goodness-of-fit is a measure of how well the best-fit template matches the recorded image. It can be used both as a check of the reconstruction quality and for signal selection.

\subsection{Signal Selection and Direct Cherenkov Light}
\begin{figure*}[htb]
\subfloat[]
	[Integrated charge per pixel.]
	{\includegraphics[width=0.32\textwidth]{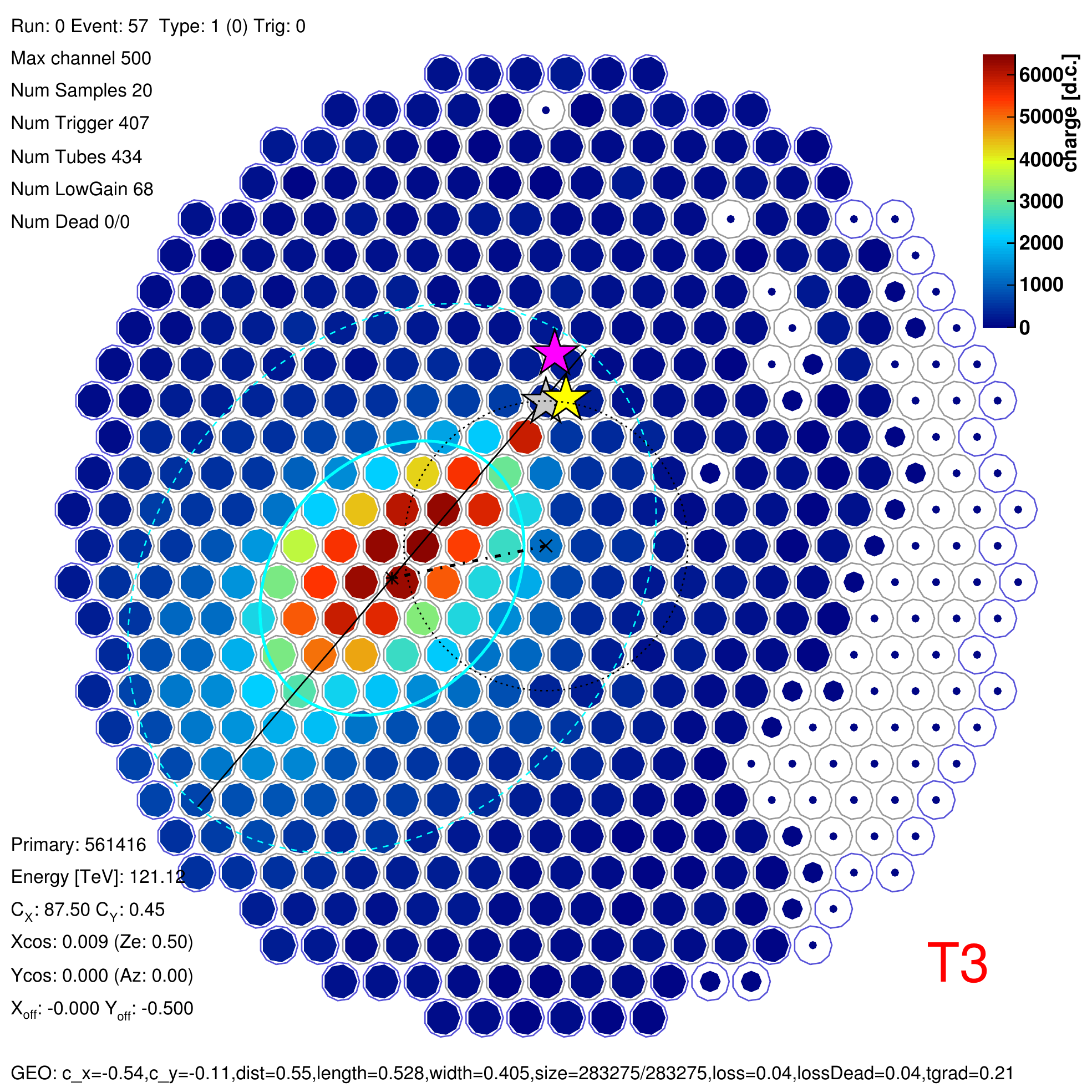}}\hfill
\subfloat[]
	[Best-fit image template.]
	{\includegraphics[width=0.32\textwidth]{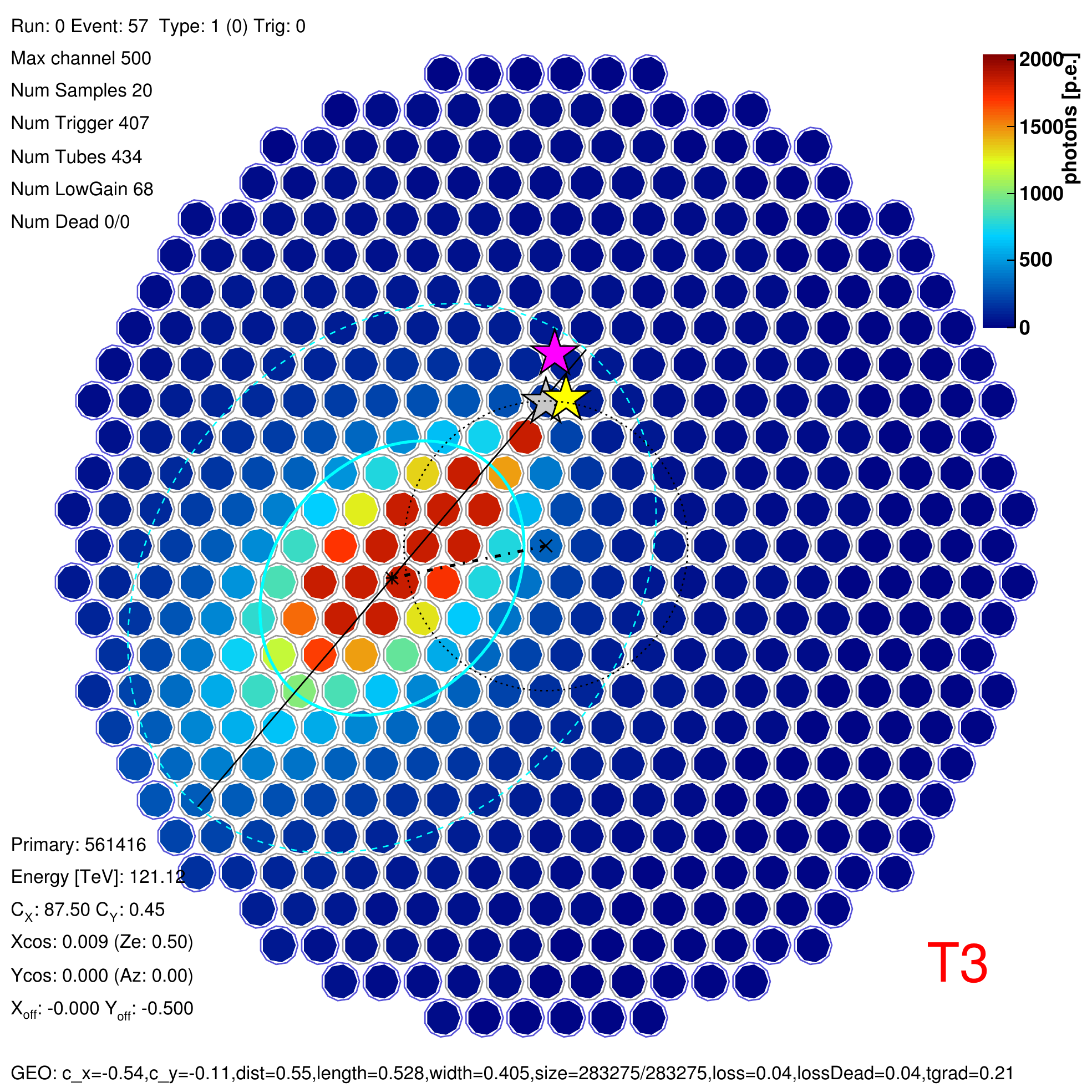}}\hfill
\subfloat[]
	[DC quality factor for the simulated image.]
	{\includegraphics[width=0.32\textwidth]{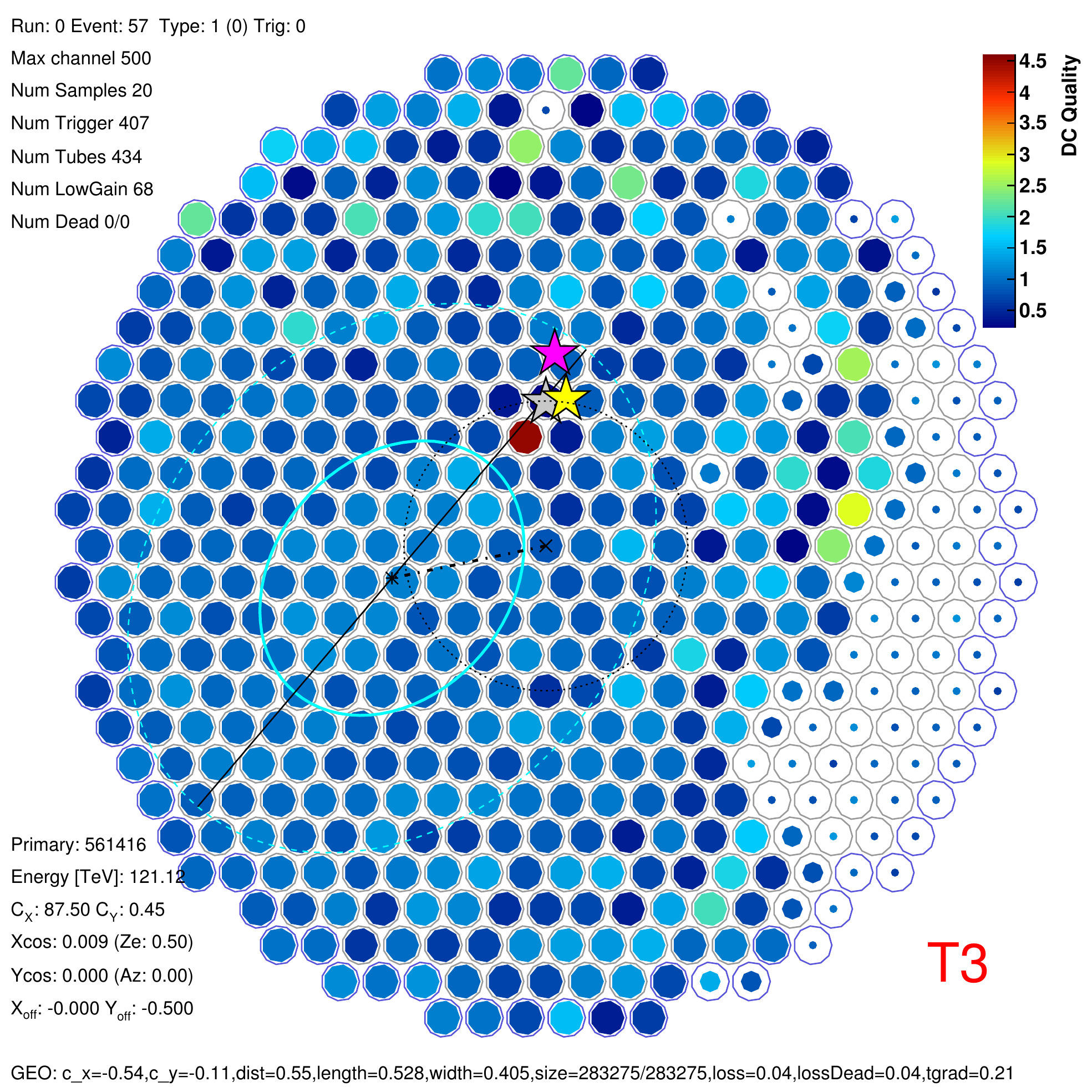}}
\caption[Simulated iron shower image]{Simulated \SI{121}{\TeV} iron shower image in one of the VERITAS telescopes. There is a visible contribution from DC light. Gray star: true shower direction. Pink star: shower direction reconstructed by the moment analysis. Yellow star: shower direction reconstructed by the template analysis. The white pixels do not contain a significant amount of signal on top of the pedestal.}
\label{fig:ironincamera}
\end{figure*}

To suppress the background due to lighter elements, only events with a DC pixel in at least two out of the four cameras were kept. This requirement provides a good compromise between efficient signal selection and background suppression. 

To identify pixels containing a contribution from the DC emission, we looked for pixels with a larger signal than their neighbors, located close to the reconstructed direction of the primary particle. For each camera, the DC pixel candidate is defined as the pixel which maximizes the DC quality factor, defined as follows (cf. \cite*{hess, wissel}): 

\begin{align}
Q_{DC,i} = \frac{q_{i}}{\left\langle q \right\rangle_{neighbors,i}},
\end{align}
where $q_i$ is the measured charge in pixel $i$ and $\left\langle \ldots \right\rangle_{neighbors,i}$ denotes the average over the neighboring pixels of pixel $i$.

In principle, the location of the DC pixel is fixed given the event geometry and the primary particle's energy. However, due to the uncertainty in the direction and energy reconstruction, a search for the DC pixel candidate is conducted over a region of the camera which fulfills the following requirements, similar to those used in \cite*{hess,wissel}: Distance to the reconstructed direction less than \ang{0.45}, distance to the image centroid between \ang{0.17} and \ang{1.2}, distance to the axis connecting the centroid to the reconstructed direction less than \ang{0.2}. 

The same search is conducted in the best-fit template image. Only images in which the DC pixel candidate in the recorded image is the same as the one in the template are kept. This has the advantage of cutting down on false positives due to statistical fluctuations, as well as removing badly reconstructed events. \Cref{fig:ironincamera} shows a simulated iron image with a noticeable DC contribution, the best-fit template image, and the value of $q_{DC}$ for each pixel. 

The contribution of DC light to the total charge in pixel $i$ can be estimated as 

\begin{align}
q_{DC,i}=q_{i}-\left\langle q \right\rangle_{neighbors,i}.
\end{align}

This may under-estimate the DC light in cases where the DC contribution is split over more than one pixel. Only DC pixel candidates with a DC contribution of at least 400 d.c. are kept. 

Following \cite{rolfdiplom} and neglecting the dependence on $\Delta h$, the \emph{reconstructed charge} in arbitrary units can be defined as

\begin{align}
Z_\mathrm{reco} = \frac{\sqrt{q_{DC}\cdot D}}{\sin \Delta_{DC}^{dir}},
\end{align}
where $D$ is the impact distance (distance between shower core and telescope) and $\Delta_{DC}^{dir}$ is the angle between the reconstructed direction of the primary and the DC candidate pixel.

For events containing more than one image with a DC candidate pixel, an average over all contributing cameras, $\left\langle Z_\mathrm{reco}\right\rangle$, is used instead. The reconstructed charge defined in this way has quite a broad distribution even for iron showers due to the aforementioned issues with over-subtraction of DC light as well as the neglected $\Delta h$ dependence. It is not suitable as a signal/background separator by itself, but may be used as an input for a multi-variate classifier. 

It must be noted that not all images of iron showers contain a visible contribution from DC light. This can be seen from the examples in \cref{fig:template}. For showers with a small impact distance, the DC light is emitted very high up, in thin air, and hence with lower intensity than for larger impact distances. Also, the centroid of the shower image and the DC light tend to overlap, making it hard to separate the two contributions. On the other hand, the DC light pool on the ground has a radius of about \SI{140}{\metre} for the VERITAS site. Showers with larger impact distances may be observed in the camera, but their images will not have a contribution from DC light. In this study, only showers with impact distances from \SIrange{40}{140}{\metre} were considered in the search for DC pixels. 

Additionally, due to the finite optical PSF of the instrument, there may be some `leakage' of DC light into one or more of the neighboring pixels. In the most extreme cases,  the DC light can be spread equally over a cluster of three connecting pixels. In that case, the DC contribution cannot be found at all by the simple algorithm employed in this study.

\subsection{Background Subtraction}
\label{sec:rf}
\begin{figure}[tb]
\includegraphics[width=0.48\textwidth]{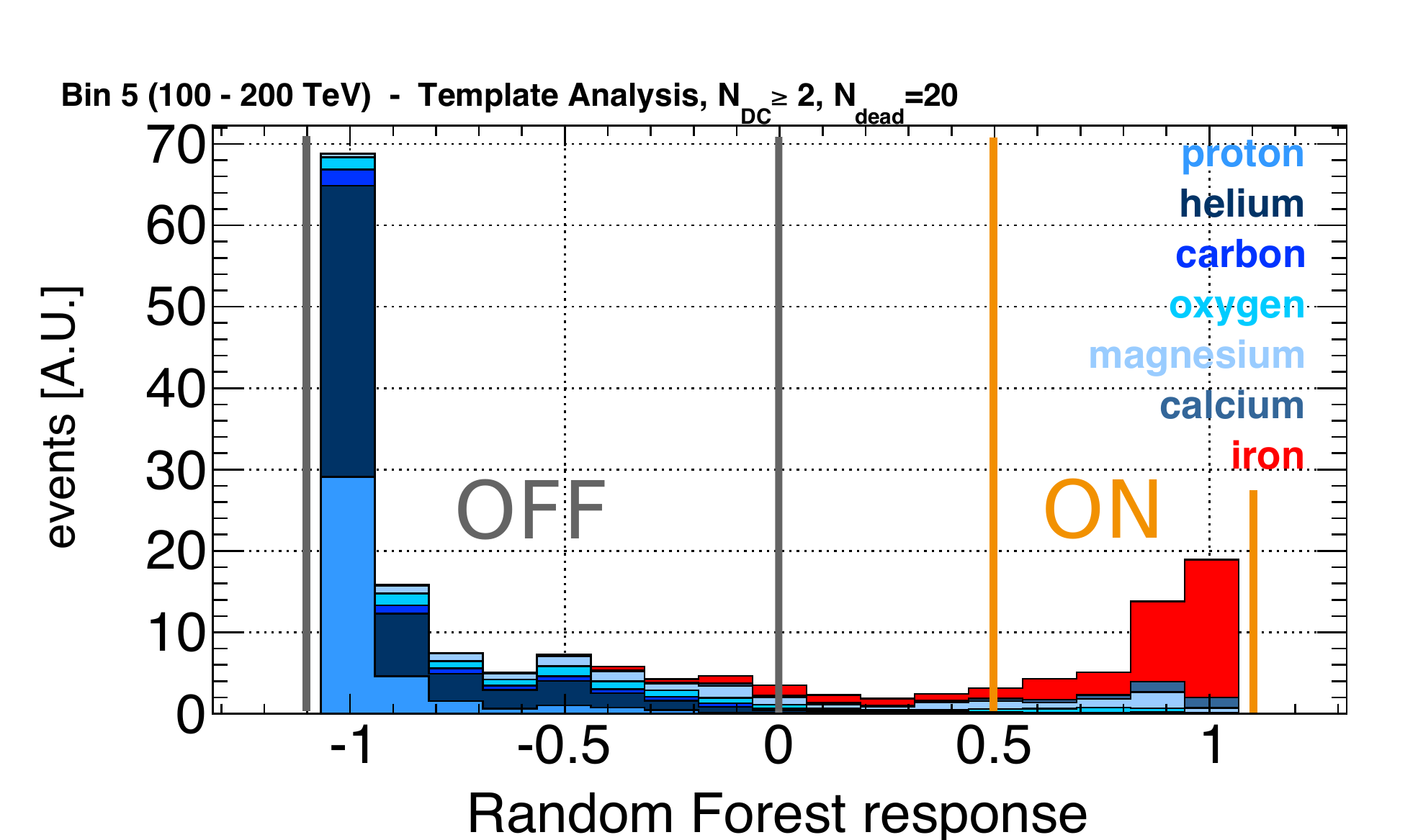}
\caption[Estimation of Signal Content Using Random Forest Response]{Response of the random forest classifier (stacked histograms) in the first energy bin from \SIrange{100}{200}{\TeV}. \emph{ON} and \emph{OFF} regions have been marked.}
\label{fig:OnOff}
\end{figure}

\begin{table*}[tb]
\centering
\sisetup{scientific-notation=true, zero-decimal-to-integer=false}
\begin{tabular*}{14cm}{@{\extracolsep{\fill} } lSSlr}
\toprule
Element range & {Normalization $f_{100}$} & {Index} & Representative &  \\
$Z$				&   {(\si{\smmr\per\steradian\per\TeV})}   	& {$\gamma$}& Element & $Z$ 		 \\
\midrule
 $1$	& \num{3.55E-07}	&\num{2.66}& H	& 1	 \\
$2-4$	& \num{4.58E-07} &	\num{2.58}	& He	& 2	 \\
$5-7$	& \num{6.04E-08}  &	\num{2.66}	& C	& 6	 \\
$8-11$ & \num{9.77E-08}	& 	\num{2.68}& O	& 8	 \\
$12-19$ & \num{1.00e-7}	& 	\num{2.64} & Mg	& 12 \\
$20-24$ & \num{2.40E-08}	& 	\num{2.70} & Ca	& 20 \\\midrule
$25-28$ & \num{1.61E-07} & \num{2.59}	& Fe	& 26  \\
\bottomrule
\end{tabular*} 
\caption[Energy spectra for simulations]{Energy spectra for simulated signal and background samples. Simulations are performed for representative elements only and weighted to a power-law spectrum $\frac{\de N}{\de E} = f_{100} \cdot 
\left( \frac{E}{\SI[scientific-notation=false]{100}{\TeV}} \right)^{-\gamma}$, where the normalization $f_{100}$ is the sum of the normalizations of the elements in the represented range and the index $\gamma$ is the index associated with the representative element. Proton and Helium spectra from \citep{0004-637X-728-2-122}, Heavier elemental spectra from \citep{WiebelSooth:1997yc}, extrapolated to higher energies.}
\label{tab:crfluxes}
\end{table*}

The number of DC pixels does not have sufficient separation power to use it as the only classifier. As mentioned before, simulations show that there is significant contamination from lighter elements even when requiring two or more images with a DC contribution. The cosmic ray flux is (to first order) isotropic, so there is no `off-source' region to estimate the background flux as can be done for the study of gamma-ray point and slightly extended sources \cite{2007A&A...466.1219B}. Using simulations to estimate the absolute normalization of the remaining background would introduce large systematic uncertainties. 

Instead, a set of \emph{random forest (RF) classifiers} \cite*{RandomForests,Albert:2007yd} was used to estimate the contributions from signal (iron) and background (lighter elements) to the signal-enriched sample of events with at least two DC pixels. Random forests are sets of decision trees, each trained on a different random subset of the full training sample. The classifiers were trained in five energy bins separately, on 15 input parameters such as the goodness-of-fit from the  template reconstruction, the mean reduced scaled length and width of the images, and the amount of DC light. 

The training samples consisted of simulated cosmic-ray air showers, including a full simulation of the VERITAS detector and readout electronics with the GrOptics and CARE packages\footnote{\protect\url{http://otte.gatech.edu/care/tutorial/}}. The elemental spectra were weighted according to previous measurements of the cosmic-ray composition, c.f. \cref{tab:crfluxes}. For proton and Helium spectra, CREAM measurements in the energy range from \SI{2.5}{\TeV} to \SI{250}{\TeV} were used \citep{0004-637X-728-2-122}. For the other elements, the results from \citep{WiebelSooth:1997yc} were used. The authors of \citep{WiebelSooth:1997yc} fitted each single element spectrum with a power law with free index and normalization, combining multiple datasets by several different direct detection experiments in the energy range of tens or hundreds of \si{GeV} to some \si{TeV}. These results were extrapolated to higher energies, assuming a constant spectral index. Only seven elements were simulated, each representing a group of elements. The spectral index obtained by \citep{WiebelSooth:1997yc,0004-637X-728-2-122} for the representative element was used for the entire group, and the flux normalization of the representative element was given by the sum of the fluxes of the group members.

The method described here does not rely on the absolute normalization of the background simulations, but instead measures both the signal and background normalization from data. It is similar to the ``template background method'' \cite{2003A&A...410..389R}, which was developed for the study of extended gamma-ray sources with IACTs. There is a remaining dependency on the assumed composition of the background sample, which is accounted for as part of the systematic uncertainty. 

The response of a random forest classifier is determined as a weighted average over the response of each decision tree, where a ``signal'' classification corresponds to $1$ and a ``background'' classification counts as $-1$. \cref{fig:OnOff} shows the response of one of the previously described RF classifiers, evaluated on an independent sample of simulated events. The distribution of the RF response peaks at 1 for signal and at -1 for background events as expected. However, the distributions are quite broad; in particular there is some irreducible background mostly due to magnesium and calcium, which emit DC light with a similar intensity to iron. 

The contributions from signal and background events to the data sample were estimated using the response of the random forest classifier. First, two disjoint intervals (ON and OFF) were defined in the random forest response in each energy bin (see \cref{fig:OnOff}). For this study, the ON interval was defined as all events with RF response of 0.5 and above (signal-dominated) and the OFF interval was defined as all events with RF response of 0 and less (background-dominated). 

In the following equations, $S$ refers to the number of signal events, $B$ to the number of background events, and $N$ to the total number of events. The subscript $_{on}$ refers to the ON region and $_{off}$ refers to the OFF region. No subscript means no cut on the random forest response was applied. The superscript $^{MC}$ indicates that the count is evaluated on simulations, while no superscript indicates that it is evaluated on data. For example, $S^{MC}$ refers to the number of simulated signal events, whereas $N_{on}$ refers to the number of data events in the ON region. 

The ON ratio, $\alpha$, and the OFF ratio, $\beta$, for signal and background, defined as follows, are obtained from simulations:

\begin{align}
\alpha_s &= \frac{S_{on}^{MC}}{S^{MC}} 
& \alpha_b &= \frac{B_{on}^{MC}}{B^{MC}} \nonumber \\
\beta_s &= \frac{S_{off}^{MC}}{S^{MC}} 
& \beta_b &=   \frac{B_{off}^{MC}}{B^{MC}} \label{eq:alphabeta}
\end{align}

Now, $N$ is the number of events in the data sample which pass the template reconstruction and have DC pixels in at least two images. This sample consists of signal and background events: $N=S+B$. The number of data events in the ON and OFF regions are then given as 

\begin{align}
N_{on} &= S_{on}+B_{on} &= \alpha_s \cdot S + \alpha_b \cdot B \nonumber\\
N_{off}&= S_{off}+B_{off}  &= \beta_s \cdot S + \beta_b \cdot B. \label{eq:onoff}
\end{align}

If this system of two equations is linearly independent (i.e. $\alpha_s\cdot \beta_b - \alpha_b\cdot \beta_s \neq 0$), it can be solved for the number of signal events:

\begin{align}
S = \frac{1}{\beta_s} \cdot \frac{N_{on} - \frac{\alpha_b}{\beta_b}\cdot N_{off}}{\frac{\alpha_s}{\beta_s} - \frac{\alpha_b}{\beta_b}}
\end{align}

$N_{on}$ and $N_{off}$ follow Poissonian distributions. For large numbers of events, these can be approximated Gaussian distributions with widths $\Delta N_{on} = \sqrt{N_{on}}$ and $\Delta N_{off} = \sqrt{N_{off}}$, respectively. The statistical uncertainty on the number of signal events is then

\begin{align}
\Delta S &= \sqrt{\left( \frac{\partial S}{\partial N_{on}} \cdot \Delta N_{on} \right)^2 + \left( \frac{\partial S}{\partial N_{off}} \cdot \Delta N_{off} \right)^2 } \nonumber \\
	&=  \left( \frac{\beta_b}{\alpha_s\cdot \beta_b - \alpha_b\cdot \beta_s} \right) \cdot \sqrt{N_{on} + \frac{\alpha_b^2}{\beta_b^2}N_{off}}.
\end{align}

\subsection{Energy Bias and Resolution}

\begin{figure*}[tb]
\subfloat[][Median energy bias (ratio of the difference between true and reconstructed energy to true energy).]{\includegraphics[width=0.48\textwidth]{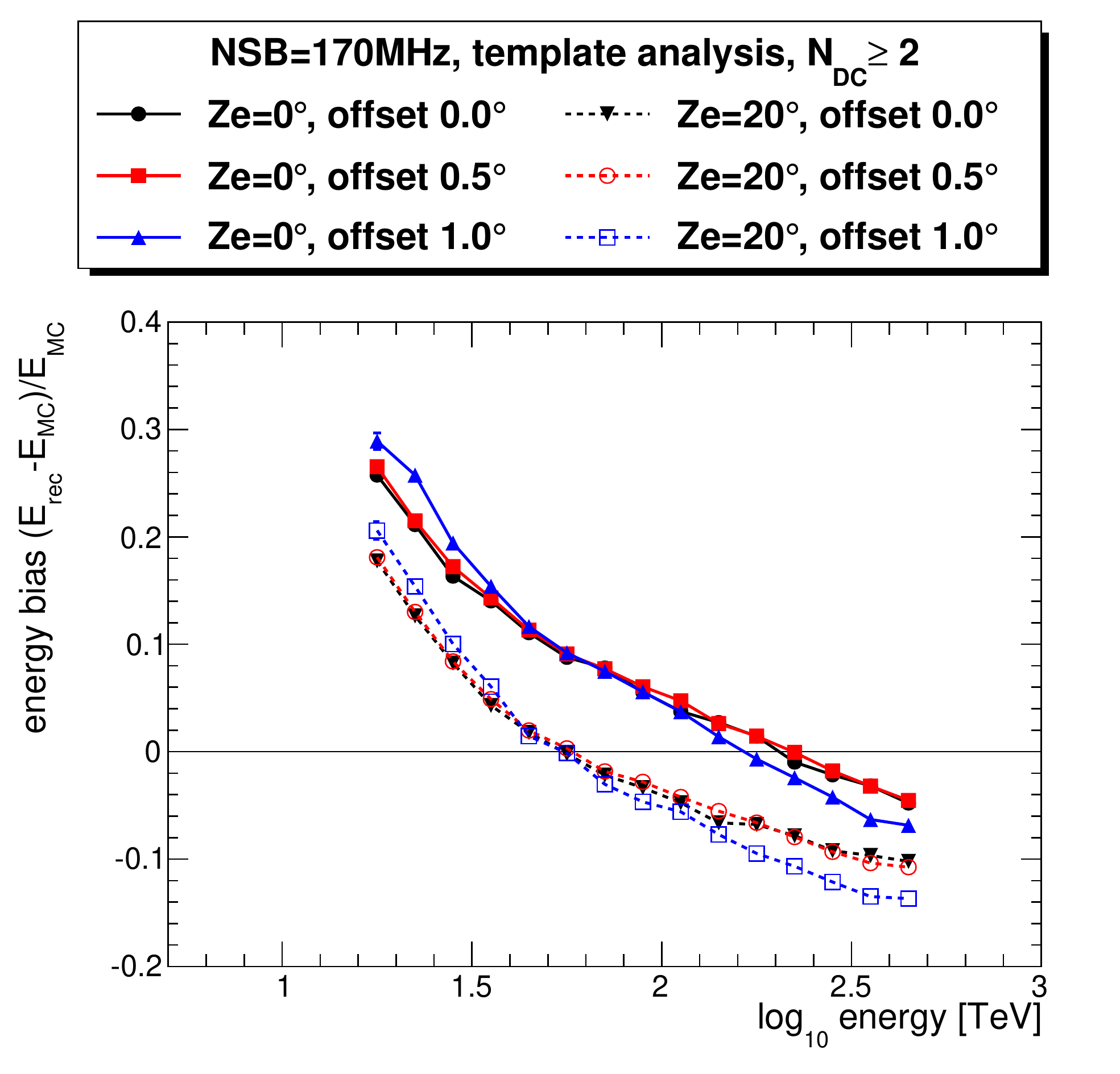}}\hfill
\subfloat[][Relative energy resolution (\SI{68}{\percent} containment interval around median).]{\includegraphics[width=0.48\textwidth]{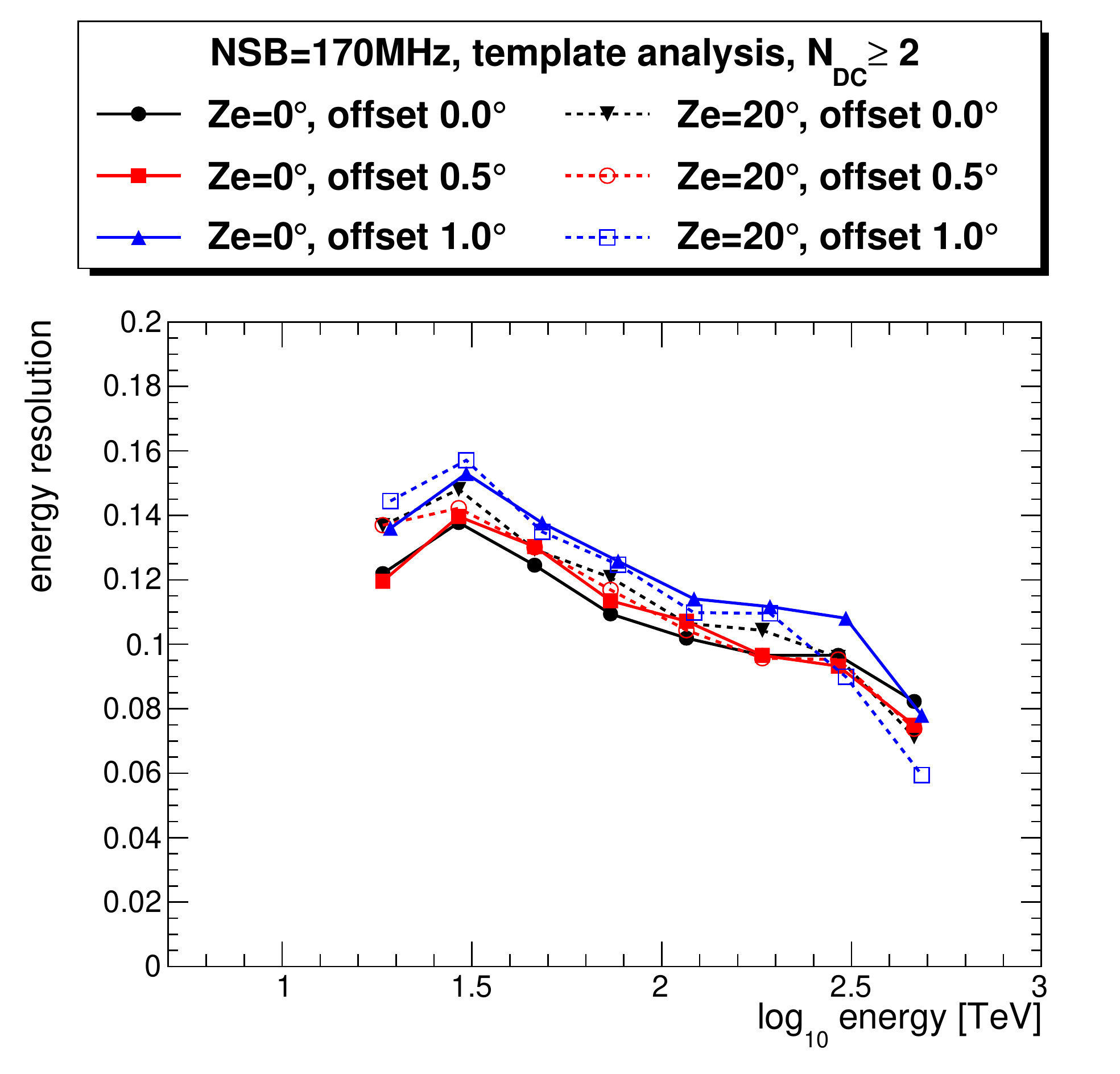}} 
\caption[Energy Bias and Resolution]{Intrinsic energy bias and resolution of the template likelihood analysis method, evaluated on simulated iron showers.}
\label{fig:EnReco}
\end{figure*}

The performance of the template likelihood reconstruction was tested on a sample of simulated iron-induced showers. The energy reconstruction is of particular interest for this study. \Cref{fig:EnReco} shows the median energy bias (ratio of the difference between true and reconstructed energy to true energy) and relative energy resolution (\SI{68}{\percent} containment interval around the median) for events with at least two images with DC pixels, depending on the zenith angle of the simulated shower as well as the offset from the camera center. 

The energy bias deviates from zero. It depends on the true energy as well as the zenith angle, and to a small extent on the offset from the camera center. Generally, the energy tends to be reconstructed about \SIrange{20}{30}{\percent} too high in the lowest energy bin (\SIrange{20}{25}{\TeV}) and up to \SI{15}{\percent} too low in the highest energy bin (\SIrange{200}{500}{\TeV}). The \SI{10}{\percent} difference in the energy bias between showers with \ang{0} and \ang{20} zenith angle is due to the fact that the templates were only produced for showers from zenith. Showers with a non-zero zenith angle produce slightly less light and are thus reconstructed at a lower energy. The energy-dependence of the energy bias is more worrying since it may cause the spectral shape to be mis-reconstructed. We believe the energy dependence of the energy bias is a result of the following selection effects: At the lowest energies, showers that produce less light than the median are less likely to pass the 70-pixel requirement. At the highest energies, showers with more light than the median are more likely to saturate the camera and cause the moment analysis to fail to produce proper starting values for the likelihood fit. Thus, the final sample contains an excess of events with $E_{reco} > E_{MC}$ at low energies and an excess of events with $E_{reco} < E_{MC}$ at high energies. The effect of this energy-dependent energy bias on the reconstructed spectrum was tested with a dedicated study \citep{myThesis}, which found that it lowered the spectral normalization by up to \SI{30}{\percent} and the spectral index by up to $0.1$. This was taken into account as part of the systematic uncertainties described in \cref{sec:systunc}. 

The energy resolution depends only weakly on the zenith angle or the offset from the camera center. It is well below \SI{20}{percent} for all energies considered here and thus already better than the typical energy resolution reached by balloon experiments (cf. \cref{sec:ironothers}).

\section{Results}

\subsection{The Cosmic Ray Iron Spectrum}

\begin{table*}[tb]
\caption[Differential Flux of Cosmic Ray Iron Nuclei]{Number of iron events and differential flux in each energy bin. $E_c$ corresponds to the logarithmic bin center. The $\alpha$ and $\beta$ ratios are defined in \cref{eq:alphabeta}. $N$ refers to the total number of data events surviving the analysis cuts. $N_{on}$ and $N_{off}$ are the number of counts in the ON and OFF regions (see \cref{fig:OnOff}), respectively, and $S$ is the derived number of signal (iron) events.}
\label{tab:ironspectrum}

\begin{tabular*}{17.8cm}{@{\extracolsep{\fill} } lrrrrrrrrrrrr}
\toprule

Bin & $E_{min}$ & $E_{max}$ & $E_c$ & $\alpha_s$ & $\alpha_b$  & $\beta_s$ & $\beta_b$& $N$ & $N_{on}$ & $N_{off}$ & $S$~~ & differential flux \\ 
  & [\si{\TeV}] & [\si{\TeV}] & [\si{\TeV}] & & & & &  &   &  & & [\si{\per\metre\squared\per\second\per\TeV\per\steradian}] \\ 
\midrule 
0 & 20 & 25.1	& 22.4 & 0.519 & 0.130 & 0.382 & 0.784 & 192 & 75 &  102 & \num[separate-uncertainty=true]{127 \pm 19}\phantom{.} & \num[scientific-notation=true, separate-uncertainty=true]{5.8 \pm  0.9 e-06} \\
1 & 25.1 & 31.6	& 28.2  & 0.658 & 0.173 & 0.263 & 0.742 & 189 & 76 & 105 & \num[separate-uncertainty=true]{86\pm 15}\phantom{.} & \num[scientific-notation=true, separate-uncertainty=true]{2.1 \pm 0.4 e-06} \\
2 & 31.6 & 39.8	& 35.5 & 0.597 & 0.113 & 0.258 & 0.807 & 171 & 51 & 103 & \num[separate-uncertainty=true]{65 \pm 13}\phantom{.} & \num[scientific-notation=true, separate-uncertainty=true, zero-decimal-to-integer=false]{1.0 \pm 0.2 e-06} \\
3 & 39.8 & 50.1	& 44.7 & 0.640 & 0.102 & 0.243 & 0.845 & 147 & 45 & 95 & \num[separate-uncertainty=true]{55 \pm 11}\phantom{.} & \num[scientific-notation=true, separate-uncertainty=true]{6.1 \pm 1.2 e-07} \\
4 & 50.1 & 100.	& 70.8 & 0.708 & 0.0644 & 0.0841 & 0.862& 337 & 67 & 229 & \num[separate-uncertainty=true]{71 \pm 12}\phantom{.} & \num[scientific-notation=true, separate-uncertainty=true]{1.5 \pm 0.3 e-07} \\
5 & 100. & 200.	& 141.3 & 0.838 & 0.0646 & 0.0798 & 0.907 & 197 & 41 & 141 & \num[separate-uncertainty=true]{37 \pm 8}\phantom{.0} & \num[scientific-notation=true, separate-uncertainty=true, zero-decimal-to-integer=false]{4.0 \pm 0.8 e-08} \\
6 & 200. & 500. & 316.2 & 0.842 & 0.0987 & 0.0722 & 0.831 & 65 & 13 & 48 & \num[separate-uncertainty=true]{8.8 \pm 4.4} & \num[scientific-notation=true, separate-uncertainty=true]{3.1 \pm 1.6 e-09} \\

\bottomrule
\end{tabular*} 

\end{table*}

\begin{figure*}[bt]
\subfloat[][Differential flux.]
{\includegraphics[width=0.48\textwidth]{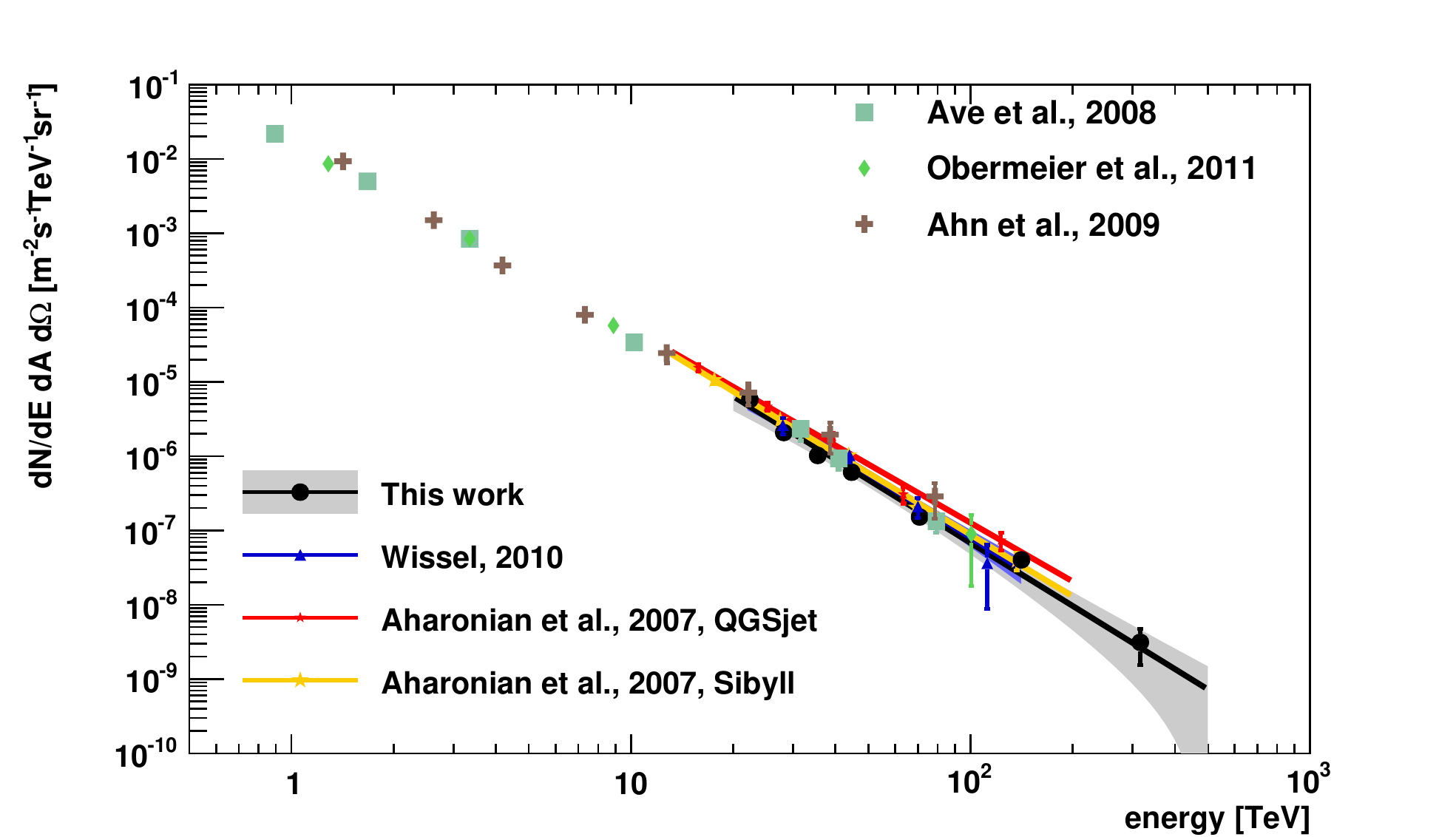}}\hfill
\subfloat[][Differential flux multiplied by $E^{2.5}$ to improve the visual clarity.]
{\includegraphics[width=0.48\textwidth]{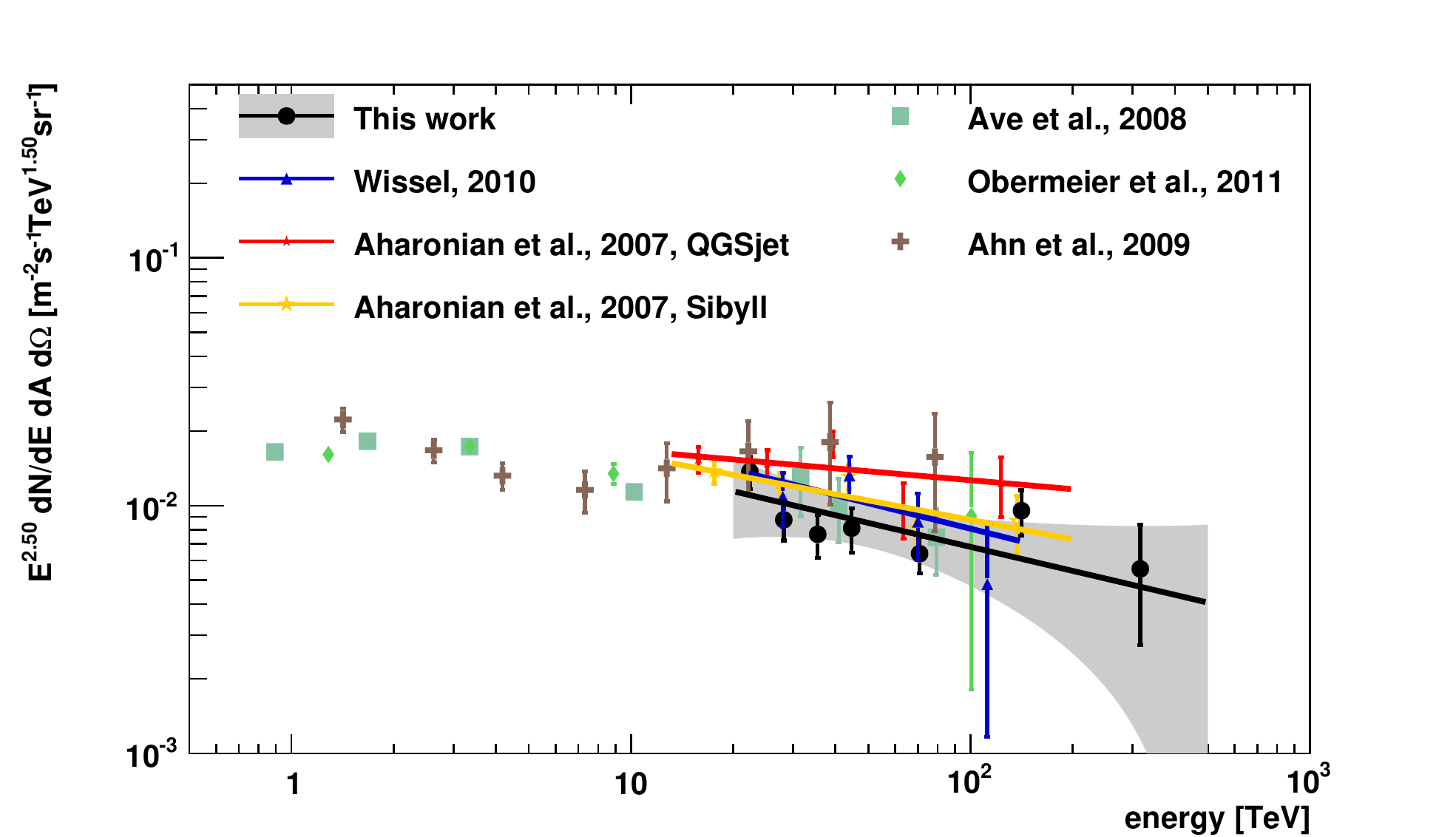}}
\caption[The Cosmic-Ray Iron Spectrum]{The energy spectrum of cosmic-ray iron nuclei as measured by the VERITAS experiment compared to previous measurements by VERITAS \citep{wissel}, H.E.S.S. \citep{hess}, and the balloon-borne detectors TRACER \cite*{Ave:2008as,2011ApJ...742...14O} and CREAM \citep{0004-637X-707-1-593}. Only statistical uncertainties are shown.}
\label{fig:ironspectrum}
\end{figure*}

The VERITAS data, selected according to \cref{sec:dataselection}, were reconstructed via the template likelihood method. The random forest classifiers were applied to the events passing the analysis cuts. Events were binned in energy: steps of 0.1 in $\log(E)$ from \SIrange{20}{50.1}{\TeV}, steps of 0.3 in $\log(E)$ from \SIrange{50.1}{200}{\TeV}, and a step of 0.4 in $\log(E)$ for the last energy bin \SIrange{200}{500}{\TeV}. The bin spacing was increased at higher energies to obtain roughly the same number of ON events in each energy bin, except for the last energy bin.

The total number of events per energy bin can be found in \Cref{tab:ironspectrum} as well as the estimated number of iron events. From this, the differential flux was calculated for each energy bin and the spectral points are plotted in \cref{fig:ironspectrum}. The resulting spectrum is well-fit by a power law

\begin{align}
\frac{\de N}{\de E \de A \de t \de \Omega} = f_0\cdot \left(\frac{E}{E_0}\right)^{-\gamma}
\label{eq:PL}
\end{align}
with normalization energy $E_0=\SI{50}{\TeV}$ over the whole energy range. The best-fit parameter values and their statistical uncertainties are: $$f_0=\SI[separate-uncertainty=true, exponent-to-prefix=false, scientific-notation=true]{4.82 \pm 0.98 e-07}{\per\metre\squared\per\second\per\TeV\per\steradian}$$ and $$\gamma=\num[separate-uncertainty=true, exponent-to-prefix=false, scientific-notation=true]{2.82 \pm 0.30}.$$

\subsection{Systematic Uncertainties}
\label{sec:systunc}
\begin{table*}[tb]
\caption[Systematic Uncertainties of the Cosmic-Ray Iron Spectral Parameters]{Systematic uncertainties of the cosmic-ray iron spectral parameters}
\label{tblSystComb}
\begin{tabular*}{17.8cm}{@{\extracolsep{\fill} } lrr}
\toprule

 Cause &  Effect on $f_0$  &  Effect on $\gamma$  \\
  \midrule
 Absolute calibration (includes atmosphere and detector model)  & $\pm \SI{40}{\percent}$ & $\pm 0.2$ \\
 `Dead' pixels (broken or turned off due to starlight)  & $\pm \SI{7}{\percent}$ & $\pm 0.07$ \\
 Intrinsic energy bias  & $^{+\SI{0}{\percent}}_{-\SI{30}{\percent}}$ & $^{+0.0}_{-0.1}$ \\ 
 Statistical uncertainty on effective area & $\pm \SI{10}{\percent}$ & --- \\ 
 Hadronic interaction model & $\pm \SI{12}{\percent}$ & $\pm 0.1$  \\
 Background sample composition & $^{+\SI{8}{\percent}}_{-\SI{13}{\percent}}$ &  $^{+0.06}_{-0.09}$ \\
 Remaining background & $^{+\SI{0}{\percent}}_{-\SI{15}{\percent}}$ & ---  \\
 \midrule
  Total & $^{+\SI{45}{\percent}}_{-\SI{56}{\percent}}$ &  $^{+0.24}_{-0.27}$\\
 \bottomrule
\end{tabular*} 
\end{table*}

The VERITAS collaboration typically assigns a \SI{20}{\percent} error to the absolute energy scale, due to the uncertainty on the absolute telescope throughput. This uncertainty has contributions from the variations in the atmospheric density and aerosol profiles, changes in mirror reflectivity, as well as uncertainties in the PMTs' quantum efficiency and absolute gain. The value of \SI{20}{\percent} was originally determined for gamma-ray showers, using dedicated air shower simulations and the relation between image \emph{size} and reconstructed energy. It is assumed to also hold for iron-induced showers as the same components contribute to the systematic uncertainty on the energy scale. For the spectral index measured here, this corresponds to an uncertainty of about \SI{40}{\percent} on the normalization. Additionally, an uncertainty of $0.2$ is assigned on the spectral index. 

Even during nominal observations, some of the PMTs may be broken, or switched off due to starlight causing the current in that pixel to exceed the safe operations limit imposed by VERITAS. Additionally, noisy pixels (either due to hardware problems or due to noise due to starlight) are not taken into account for the analysis. The selection of DC pixels is particularly sensitive to these `dead' pixels. In the dataset considered for this study, there were on average 20 dead pixels in each camera. This was taken into account for the detector simulations. To assess the effect of the number of dead pixels, the simulations were repeated with 0 and 40 dead pixels per camera, bracketing the distribution in data. The effect on the reconstructed spectrum was very small. Accordingly, an uncertainty of \SI{7}{\percent} was assigned to the normalization and an uncertainty of 0.07 was assigned to the spectral index due to the effect of dead pixels. 

The template reconstruction method has an intrinsic energy bias, related to not fitting the first interaction height as well as the selection of DC candidates. This energy bias, evaluated with simulations, varies with energy and thus affects the shape of the measured spectrum. It also depends on the zenith angle of the shower. A dedicated study \citep{myThesis} was conducted to estimate the effect of this energy bias and accordingly, an uncertainty of $^{+\SI{0}{\percent}}_{\SI{-30}{\percent}}$ was assigned to the normalization and an uncertainty of $^{+0}_{-0.1}$ was assigned to the spectral index. 

To assess the dependence on the hadronic interaction model, additional showers were simulated using the Sybil 2.1 interaction model \cite{2009PhRvD..80i4003A}. Uncertainties of \SI{12}{\percent} on the normalization and 0.1 on the spectral index were found.

An additional \SI{10}{\percent} uncertainty on the normalization was found due to the statistical uncertainty on the effective area.

The systematic uncertainty related to the composition of the background sample was evaluated in two different ways. First, the fluxes and spectral indices of the elements were varied randomly according to the quoted uncertainty, assuming gaussian uncertainties and zero correlation. The calculation of the ON/OFF ratios and flux points and the spectral fit were repeated for each random background composition. The resulting distributions of the flux normalization and index were significantly narrower than the statistical uncertainties on the flux and index, thus this contribution was neglected here. In a second study, the flux/spectral index of each element was separately shifted up and down. For protons and helium, the flux normalization was shifted up/down by a factor of two. For the heavier elements, the index was shifted by $\pm 0.15$ while keeping the flux normalization at \SI{1}{\TeV} fixed, corresponding to a shift in the flux norm at \SI{100}{\TeV} by a factor of 2. This is considered the worst-case scenario given recent measurements of elemental spectra in the \si{TeV} range. The measured iron flux normalization is most sensitive to a change in the assumed Ca group spectrum. The measured iron spectral index is most sensitive to the change in the assumed proton (+0.06/-0.07) and Mg group (+0.05/-0.09) spectra. We adopt those maximal changes in the measured iron spectrum as systematic uncertainties, see \cref{tblSystComb}.

Finally, there is a contribution to the uncertainty due to remaining background events. Elements up to chromium ($Z=24$) were included in the background simulations. However, there could be remaining contributions from manganese ($Z=25$) and nickel ($Z=28$), which, according to \cite{Hoerandel:2002yg} have fluxes of \SI{10}{\percent} (manganese) and \SI{7}{\percent} (nickel) of the iron flux at \si{\TeV} energies. The contribution from all other elements can be neglected as their fluxes are below $\SI{1}{\percent}$ of the iron flux at \SI{1}{\TeV}. Assuming that the relative abundances do not change up to \SI{500}{\TeV} and that the random forest response to these elements is similar compared to iron, the measured `iron' flux may consist of up to \SI{15}{\percent} other elements (manganese and nickel), which is assigned as an additional uncertainty on the normalization. 

The image templates were generated for showers from zenith only, which affects the energy reconstruction (showers with non-zero zenith angle tend to be a bit dimmer). This is part of the energy bias discussed above. The random forests were also trained on showers from zenith only, as the dependence on the $\alpha$ and $\beta$ parameters (used in the background subtraction) on the zenith angle was found to be negligible. The effective area used in the calculation of the differential flux points was interpolated from simulations at \ang{0} and \ang{20} zenith angle, and thus correctly takes into account the zenith angle dependence.

The instability of the fit to the first interaction height, which caused the first interaction height to be removed from the likelihood fit parameters, slightly degrades the energy resolution, and might cause a slight energy bias. As the fit parameters were the same for data and simulations, this effect is included in the energy bias discussion above.

These contributions to the systematic uncertainty are summarized in \cref{tblSystComb}. The dominant source of uncertainty is the uncertainty on the absolute throughput due to the atmosphere and detector model. The different sources of systematic uncertainty were assumed to be uncorrelated and added in quadrature to obtain the total systematic uncertainty. The resulting final measurement of the parameters of the energy spectrum of iron nuclei in cosmic rays is:

\begin{align*}
f_0 &= ( 4.82 \pm 0.98_{stat}\,^{+2.12}_{-2.70 sys} )  \cdot \num[scientific-notation=true, exponent-to-prefix=false,retain-unity-mantissa = false]{1e-7} \si{\per\metre\squared\per\second\per\TeV\per\steradian} \\
\gamma &= 2.82 \pm 0.30_{stat}\,^{+0.24}_{-0.27 sys},
\end{align*}
where $f_0$ is the normalization at $\SI{1}{\TeV}$ and $\gamma$ is the spectral index.

\subsection{Comparison to Previous Measurements}
These results are compared with earlier measurements of the iron spectrum in \cref{fig:ironspectrum}. In particular, we would like to highlight the earlier measurement made by VERITAS \cite{wissel}. The observations analyzed in the previous study are an independent dataset from the one that was analyzed with the template method. The earlier results are based on an analysis of 397 hours of observations taken under favorable weather conditions between September 2007 and May 2009. Data selection cuts required that all four telescopes were active and pointed at mean elevation angles of \SIrange{70}{80}{\degree}. Direct Cherenkov events were selected based on shower parameters, directional reconstruction, and DC-pixel identification, similar to \cite{hess}. Iron showers were selected from the sample by reconstructing the charge of each event based on the number of Cherenkov photons and the distance between the telescope and the shower core. Due to the harder event selection, the previous analysis yielded fewer signal events (total of 57 events), but was able to suppress all light backgrounds. The final sample is estimated to contain at most \SI{14}{\percent} of heavier nuclei aside from iron. 

The energies of the iron events were reconstructed using look-up tables based on standard Hillas parameters \cite{hillas}. The tables were generated from iron shower simulations produced by CORSIKA 6.702 \cite{corsika}, using FLUKA 2006 \cite{Battistoni:2007zzb} for low-energy interactions and QGSJET-II \cite{2006NuPhS.151..143O} for the high-energy interactions. The VERITAS telescopes were modeled using the GrISU package \cite{grisu}. The resulting energy spectrum was measured in the range from \SIrange{22}{140}{\TeV} with \SI{24}{\percent} energy resolution, and was unfolded according to \cite{1994ApJ...429..736B}. The energy spectrum from the 2007--2009 dataset was found to follow a power law with normalization

\begin{align}
f_{0} = (5.8\pm0.84_{stat}\pm 1.2_{sys}) \SI[scientific-notation=true, exponent-to-prefix=false,retain-unity-mantissa = false]{1e-7} {\per\metre\squared\per\second\per\TeV\per\steradian}
\end{align}
at \SI{50}{\TeV} and index $\gamma = 2.84\pm0.30_{stat}\pm0.3_{sys}$. 

The two VERITAS measurements agree with each other within statistical uncertainties, as shown in \cref{fig:ironspectrum}. The VERITAS results are also compatible with previous measurements of the iron spectrum using direct detection \cite*{Ave:2008as,2011ApJ...742...14O,0004-637X-707-1-593} and the direct Cherenkov technique \cite*{hess} within the statistical uncertainties. In contrast to previous work using imaging Cherenkov telescopes to measure the iron spectrum \cite{hess,wissel}, the template-based analysis extends the spectrum up to \SI{500}{\TeV} and provides smaller statistical uncertainties above \SI{50}{\TeV}. 

The energy threshold for the template-based analysis is \SI{20}{\TeV}, which is slightly larger than for the moment-based H.E.S.S. analysis \cite{hess}. This is most likely due to the strict quality cuts (at least 70 hit pixels in each camera) imposed here, chosen to ensure good performance of the template fit rather than to optimize the sensitivity at low energies.

\section{Future Outlook}
Future iterations of this analysis can build on the present results in several ways. For example, a successful fit of the first interaction height would enable the measurement of the iron-air cross-section and the test of hadronic/nuclear interaction models, independent of accelerator experiments. Additionally, the successful fit of the first interaction height is expected to further improve the energy resolution. If the uncertainty on the absolute energy scale can also be improved, future studies will potentially be sensitive to features in the intrinsic spectrum, for example any spectral hardening. For instance, the Cherenkov Telescope Array (CTA) \cite{2013APh....43....3A}, a next-generation gamma-ray observatory currently under development, will be a good candidate to conduct such studies. It will have a larger collection area and improved energy resolution compared to current instruments, and with its two sites, each comprising tens of Cherenkov telescopes of varying sizes, it might also be able to cover a larger energy range. 

Even though this analysis was optimized for the selection of iron-induced showers, it can already be seen that the response of the random forest classifier is different for proton/helium-induced showers and showers induced by intermediate elements such as carbon and oxygen. A dedicated analysis with templates for the different elements/elemental groups should be able to measure not just the iron spectrum, but also the elemental composition in the \si{\TeV} to \si{\PeV} energy range.

Last, but not least, as the cosmic ray spectra in this energy range are not affected by solar modulation, they are expected to be constant on timescales of hours to years. As the CTA array will have more telescopes covering a larger area on the ground as well as a larger field-of-view, it should be able to collect a similar amount of statistics within a few weeks or months. Thus, measurements of the cosmic-ray iron flux could be used to test the detector stability over time. 

\section{Summary and Conclusions}
In this paper, we have presented an updated measurement of the energy spectrum of cosmic-ray iron nuclei with the VERITAS experiment in the energy range of \SIrange{20}{500}{\TeV}. Despite the remaining problems with the fit of the first interaction height, this work demonstrates the power of the template likelihood method as applied to the analysis of cosmic-ray data recorded by IACTs. 

The measured energy spectrum is compatible with a power-law shape, agreeing well with previous measurements in this energy range within the statistical and systematic uncertainties. The uncertainty is dominated by the uncertainty on the absolute energy scale due to variations in the atmosphere and detector. Due to the large uncertainties (compared to direct detection experiments), this analysis is not sensitive to a potential hardening in the spectrum of $\Delta\gamma\approx 0.1$ such as the one observed by AMS in the proton and helium spectrum. Similarly, while there is no indication for a cutoff in the spectrum given the present data, this analysis is not sensitive to a potential cutoff or softening in the iron spectrum above hundreds of TeV.

The template-based analysis has a slightly higher energy threshold than the two previous measurements of the iron spectrum with IACTs. On the other hand, the template fit reconstruction is able to compensate to some extent the loss of information which occurs when large images are not fully contained in the camera. This improved the sensitivity at the highest energies and enabled us to measure the spectrum up to \SI{500}{\TeV}. 

Future experiments such as the CTA observatory will be able to improve upon the measurements presented here, for example using the template likelihood method presented here to measure the spectra of more elements.

\begin{acknowledgments}
VERITAS is supported by grants from the U.S. Department of Energy Office of Science, the U.S. National Science Foundation and the Smithsonian Institution, and by NSERC in Canada. We acknowledge the excellent work of the technical support staff at the Fred Lawrence Whipple Observatory and at the collaborating institutions in the construction and operation of the instrument. The VERITAS Collaboration is grateful to Trevor Weekes for his seminal contributions and leadership in the field of VHE gamma-ray astrophysics, which made this study possible. H. Fleischhack gratefully acknowledges support through the Helmholtz Alliance for Astroparticle Physics.
\end{acknowledgments}

\bibliography{bibliography.bib}

\end{document}